\documentclass[aps, tightenlines, preprint, nofootinbib]{revtex4-1}

\usepackage{eurosym} 
\usepackage{amsmath} 
\usepackage{amssymb} 
\usepackage{dsfont} 
\usepackage{enumerate} 
\usepackage{color}
\usepackage{graphicx} 
\usepackage[mathscr]{eucal} 
\usepackage{longtable} 
\usepackage{multirow}

\newcommand{\dd}{\mathrm{d}}
\newcommand{\Dp}{\partial}
\newcommand{\un}{\infty}

\newcommand{\li}{\left}
\newcommand{\ri}{\right}

\newcommand{\cen}[1]{\begin{center} #1 \end{center}}

\newcommand{\blau}[1]{\textcolor{black}{#1}}

\begin{document} {\normalsize}
\title{Phase structures emerging from holography with \\ Einstein gravity - dilaton models at finite temperature}

\author{R. Z\"ollner,}
\author{B. K\"ampfer}
\affiliation{Helmholtz-Zentrum Dresden-Rossendorf, \\PF 510119, D-01314 Dresden, Germany \\ and \\ Institut f\"ur Theoretische Physik, TU Dresden, \\D-01062 Dresden, Germany}

\begin{abstract}
Asymptotic AdS Riemann space-times in five dimensions with a black brane (horizon) sourced by a fully back-reacted scalar field (dilaton) offer -- via the holographic dictionary -- various 
options for the thermodynamics of the flat four-dimensional boundary theory, uncovering Hawking-Page, first-order and second-order phase transitions up to a cross-over or featureless behavior. The 
relation of these phase structures to the dilaton potential is clarified and illustrating examples are presented. Having in mind applications to QCD we study probe vector mesons with the 
goal to figure out conditions for forming Regge type series of radial excitations and address the issue of meson melting. 
\end{abstract}

\maketitle

\section{Introduction}
The advent of the insight into AdS/CFT correspondence \cite{maldacena, witten, gubser} offered the option of having an alternative access to strongly coupled systems, e.g.~to various facets of QCD in 
the non-pertubative regime, for instance.
Phenomenologically interesting problems, e.g.~the hadron spectrum or properties of the quark-gluon plasma, become treatable within a framework called holographic approaches. Most desirably would be to 
have a holographic
QCD dual at our disposal, from which statements on QCD-related
quantities can be derived in a unique manner. However, such a dual is
presently not available \cite{dual, dual2, dual3, KKSS}. Therefore, in practice, field-theory quantities in a five dimensional asymptotic Anti-de Sitter (AdS) are often related to observables (or expectation values of operators)
in four dimensional Minkowski space-time in the spirit of the
field-operator duality (\cite{dictionary}, see also \S5.3 in \cite{erdmenger}, \S10.3 in \cite{nastase}, for instance). Top-down approaches attempt to use input
from string theory constructions – or elements thereof. These are to be
contrasted with bottom-up approaches which aim at starting with an
appropriate model on the field-theory side to mimic certain selected
features of the boundary theory side, with the latter being connected
to the quantum field theory in Minkowski space, while the former
includes the dynamics in the bulk. \\
Besides early emphasis on accessing principal features of strongly
coupled systems with many extensions to higher or lower dimensions than
mentioned above, one can also take the attitude of adjusting
sufficiently simple and thus transparent bottom-up models to a certain
input and then employ them for predictions. Of course, the predictive
power becomes a relevant issue here. Moreover, the foundations of the
AdS/CFT correspondence, namely a very large number $N_c$ of gauge
degrees of freedom and a very large 't~Hooft coupling, are often
argued only to hold under special conditions, too. For instance, w.r.t.~QCD one knows
\cite{panero} that certain thermodynamic observables of the Yang-Mills gauge theory obey the proper scaling with $N_c$ and
hopes that the physics case of $N_c=3$ is adequately captured. \\
Holographic modeling of QCD related problems became popular due to some
particularly striking findings. Among them are the Regge type spectrum
of hadronic and glue ball states, e.g.~within the soft-wall model \cite{KKSS},
the famous ratio of shear viscosity $\eta$ to entropy density $s$,
$\eta/s = 1 /4\pi$ \cite{kss}, and the phase diagram with a critical point \cite{wolfe, wolfe2}, to mention a few ones. \\
Besides gravity, the dilaton plays an important role as a breaker of the
conformal symmetry since it introduces an energy scale. Obviously,
the holographic models with gravity coupled to and sourced by a dilaton
field -- including the negative cosmological constant to ensure the
asymptotic AdS geometry -- represent some minimalistic set-up. To be
specific, we restrict ourselves here to Einstein gravity. What remains
is fixing the dilaton self-interaction. This may refer to roots in
string theory, as recently put forward, e.g.~in \cite{debye, topdown} in a top-down
approach, or to shape the dilaton self-interaction -- encoded in the
dilaton potential -- by reproducing a certain set of wanted results
within the dilaton engineering to reproduce Lattice QCD thermodynamics
results of \cite{borsa, baza}. \\
The resulting set-up is called Einstein gravity - dilaton model. It continues numerous previous studies in cosmology, most notably in
inflationary scenarios. Analogously, in holography such gravity - dilaton models enjoy some popularity due to their conceptual simplicity. There
is an overwhelming number of studies, e.g.~\cite{debye, exp1, exp2, innes, mimi, cherman, hohler, fin6, kir1, kir2, kir3, danning, knaute2, fin2, yaresko, yaresko-knaute, alanen3, alanen1, alanen2, 
fin3, fin4, fin5, braga, rouge, rouge2, pufu, rocha, hydro1, hydro2, spring, afonin, fairness, paula, paula2, paula3, mei, mei2, mei3, danning2, vega, vega2, wang, gutsche, reece, brax, rees, burg, 
cadoni2, cadoni1}, based on that type of model:
\cite{exp1, exp2} represent an in-depth analysis and review of the model in detail and \cite{innes} touches the issue of consistency.
References~\cite{debye, mimi, cherman, hohler, fin6, kir1, kir2, kir3, danning, knaute2, fin2, yaresko, yaresko-knaute, alanen3, alanen1, alanen2, fin3, fin4, fin5, braga, rouge, rouge2, pufu, rocha} 
focus on thermodynamics, where \cite{mimi} can be considered to be the prototype of modeling thermodynamics with a dual black hole. The authors of \cite{cherman, hohler} claim a bound of the 
speed of sound and \cite{fin6} derived a relation between the speed of sound and the single heavy quark free energy.
Discussions about different phase structures can be found in \cite{kir1, kir2, kir3, danning}, and for quantitative comparisons and parameter fits to results on Lattice QCD thermodynamics in case of 
vanishing and finite baryo-chemical potential  we refer to \cite{knaute2, fin2, yaresko, yaresko-knaute, alanen3}.
There are investigations about the temperature dependence and the behavior during phase transitions of related quantities, e.g.~\cite{alanen1} calculates string tension at finite temperature, 
\cite{alanen2} chooses an approach based on the beta function, \cite{fin3, fin4, fin5} deal with electric and magnetic quantities, \cite{debye, braga} calculate the Debye screening mass; transport 
coefficients and bulk viscosities are the topics of \cite{rouge, rouge2} and \cite{pufu, rocha}, respectively. A holographic approach to the broad field of hydrodynamics and thermalization is given 
e.g.~by \cite{hydro1, hydro2, spring} within the gravity - dilaton model class. \\
The soft-wall model \cite{KKSS} developed into a role model for computing particle spectra holographically. While in the original model the metric background is fixed by ad-hoc ans\"atze, the main 
idea of its generalizations \cite{afonin, fairness, paula, paula2, paula3, mei, mei2, mei3, danning2, vega, vega2, wang, gutsche} is to obtain the metric background as a solution of the Einstein 
equations. The particles are considered then as test particles. \cite{afonin, fairness, paula, paula2, paula3} give particular attention to Regge type spectra, \cite{mei2} investigates the case of 
finite temperature and \cite{mei3, danning2} focus on chiral symmetry breaking.
Due to the various applications of Einstein gravity - dilaton models (see as well \cite{reece} for a braneless approach, \cite{brax} for fluctuating branes, \cite{rees} for a real-time formulation, 
\cite{burg} for cosmological discussions, \cite{cadoni2} for scalar condensates or \cite{cadoni1} for a generalization to higher dimensions) this list does not purport to be complete.\\
Finite temperature effects are generated by plugging a black hole in the originally AdS and deform it accordingly. Thus, a Hawking temperature and a Bekenstein-Hawking entropy density link to thermodynamics.
In the present paper, we also stay within such a framework: holographic
gravity-dilaton model with the goal to elaborate the emerging
thermodynamics w.r.t.~conditions for the dilaton potential to catch
certain phase structures with relevance to QCD.\footnote{Such an investigation
is timely, since a systematic study relating the dilaton potential and
the emerging thermodynamics is currently lacking \cite{ammon}, see however \cite{hydro2}, where selected cases are considered.} The Columbia plot (cf. \cite{philipsen2018} for an updated
version) provides several options for 2+1 flavor QCD: in dependence on
the quark masses, first- and second- order phase transitions may show up
as well as a cross-over and some others. We try to answer the question which properties the dilaton potential must have to enable these phase properties
related to deconfinement and chiral restoration in QCD. On top of
thermodynamic aspects we consider holographic probe vector mesons. That is, the gravity and dilaton background resulting from
the field equations and equation of motion governs -- besides the
thermodynamic features -- the behavior of vector mesons. We thus extend
our previous studies \cite{ich2017} and investigate to which extent the
disappearance of vector mesons as a possible indicator of deconfinement occurs
at the QCD cross-over temperature. This is important for a holographic
realization of the thermo-statistical interpretation of hadron
multiplicities in ultra-relativistic heavy-ion collisions, e.g.~at LHC~\cite{alice}. The ultimate goal of such investigations, which however is beyond
the scope of the present paper, is an extension to non-zero chemical
potential, e.g.~to address issues of the QCD phase diagram and the
chemical freeze-out curve therein. \\
Our paper is organized as follows. In Section \ref{sec2}, we recall the
holographic settings, that is the gravity-dilaton model, its field
equations and equation of motion as well as the emerging
thermodynamics and the access to phase structure. Also, the holographic description of vector mesons probing the background is
recalled. Section \ref{sec3} deals with thermodynamic scenarios by presenting a
series of selected examples of transition types characterized by
entropy density, sound velocity and pressure. This is supplemented by
showing the Schr\"odinger equivalent potential which governs the
existence or non-existence of vector mesons. After some general remarks on
shaping the dilaton potential, we try to elucidate the conditions on
the dilaton potential to enforce a first-order phase transition or a
cross-over. In Section \ref{sec4}, we explain a relation between the
Schr\"odinger equivalent potential at zero temperature and the
thermodynamic features. Both ones are linked by the field equations;
details can be found in Appendices \ref{anhang_A} and \ref{anhang_B}. In the second part of
Section \ref{sec4}, we reverse our view: instead of starting with a dilaton
potential we model a certain shape of the Schr\"odinger equivalent
potential which allows for a certain wanted hadron spectrum (ideally of a Regge type) and derive --
again via field equations -- the resulting dilaton potential.\footnote{Alternatively, one could also start with an ansatz for the dilaton profile and derive all other quantities via field equations, 
cf.~\cite{mei} and further references therein. Obviously, one could start equally well with other quantities or combinations thereof and derive the remaining functions from the field equations. 
\cite{paula} is an example for starting with the warp factor, defined below.} We
conclude that part by considering the vector meson melting upon
temperature increase. The summary and a discussion of possible
extensions towards the goal of a consistent scenario of QCD
thermodynamics with the chemical freeze-out model in the LHC energy
regime can be found in Section \ref{sec5}.

\section{Holographic settings} \label{sec2}
\subsection{Thermodynamics from Einstein gravity - dilaton model} \label{sec2-1}

We consider the action\blau{\footnote{
\blau{The rational is concisely formulated in  \cite{mimi}:
"We would like to find a five-dimensional gravitational theory that has black hole solutions whose speed of sound as a function of temperature mimics that of QCD. We will not try to include 
chemical potentials or to account for chiral symmetry breaking.  We will not try to include asymptotic freedom, but instead will limit our computation to $T<4T_c$ and assume conformal behavior in the 
extreme UV. We will not even try to give an account of confinement, except insofar as the steep rise in the number of degrees of freedom near the cross-over temperature $T_c$ is recovered in our 
set-up, corresponding to a minimum of $c_s$ near $T_c$. We will not try to embed our construction in string theory, but instead adjust parameters in a five-dimensional gravitational action to recover 
approximately the dependence $c_s (T)$ found from the lattice. ... We will not include higher derivative corrections, which would arise from $\alpha'$ and loop corrections if the theory ... were 
embedded explicitly in string theory.''}}}
  \begin{equation}
   S=\frac1{2\kappa} \int \! \dd^5 x \sqrt{g}\li[ R-\frac12 (\Dp \Phi)^2-V(\Phi) \ri] \label{wirkung}
  \end{equation}
over the five-dimensional Riemann space-time with special ansatz of the metric given by the line element squared as
  \begin{equation}
   \dd s^2 = e^{A(z)} \li[f(z) \dd t^2 - \dd \vec x^2 -\frac1{f(z)} \dd z^2\ri],  \label{metrik}
  \end{equation}
where $A$ denotes the warp factor with $A(z) \to -2\ln (z/L)$ as $z \to 0$ to ensure an asymptotic Anti-de Sitter (AdS) and $f$ denotes the blackness function with $f(z)\to1-\mathcal{O}(z^4)$ as $z 
\to 0$ which encodes the temperature via
  \begin{equation}
   T(z_H)=-\frac1{4\pi} f'(z_H) \label{temp}
  \end{equation}
with the horizon position $z_H$ and the simple zero $f(z_H)=0$.
\pagebreak

The vacuum case, $T=0$, is equivalent to $f=1$. The dilaton $\Phi$ in the action \eqref{wirkung} is a dimensionless real-valued 
scalar bulk field with $\Phi \propto z^{\Delta} + z^{4-\Delta}$ if $z \to 0$, where the conformal dimension $\Delta$ as the larger solution of $\Delta (4-\Delta)=L^2 m_{\Phi}^2$ is related to the 
dilaton 
mass $m_{\Phi}$. Its potential $V(\Phi)$ has the asymptotic small-$\Phi$ form $L^2 V = -12-\frac12 L^2m_{\Phi} ^2 \Phi^2+ \cdots$, where the first term refers to the negative cosmological constant and 
the second one has to obey the Breitenlohner-Freedman (BF) bound $-4 \leq m_{\Phi} ^2L^2 \leq 0$ \cite{BF1, BF2}; $L$ sets a scale, as $\kappa$ in (\ref{wirkung}), to make the action dimensionless in 
natural units. From (\ref{wirkung}) and (\ref{metrik}) the field equations follow
  \begin{eqnarray}
   f'' +\frac32 A' f' &=& 0, \label{fgl1} \\
   A''-\frac12 {A'}^2 +\frac13 {\Phi'}^2 &=& 0, \label{fgl2} \\
   ({A'}^2-\frac16 {\Phi'}^2) f + \frac12 A'f' +\frac13 e^A V &=& 0, \label{fgl3}
  \end{eqnarray}
and the equation of motion
  \begin{equation}
   \Phi'' +\li( \frac32 A'+\frac{f'}f\ri) \Phi' -\frac{e^A}f \Dp_{\Phi} V =0 \label{fgl4}
  \end{equation}
which is redundant since it follows from (\ref{fgl3}) with (\ref{fgl1}, \ref{fgl2}). A prime means derivative w.r.t.~the bulk coordinate $z$. Equations (\ref{fgl1}-\ref{fgl3}) can be solved for a 
given $V(\Phi)$ with the above side conditions. Note that the integration constants have a higher degree of freedom in the vacuum case ($T=0$) than in 
case of finite temperature due to the side condition $f=0$ at the horizon $z_H$. That can be seen, for instance, by series expansions in various coordinates, exhaustively done in the literature, 
e.g.~\cite{wolfe, wolfe2, exp1, exp2, boschi}. The request of the continuous embedding of the vacuum quantities in the set of the finite temperature quantities for all choices of $V$ allows for 
picking up the admissible vacuum solution. \\
In such a way, the dilaton potential $V(\Phi)$ determines the temperature via (\ref{temp}) and the speed of sound squared via
  \begin{equation}
   c_s^2 = \frac{\dd \ln T}{\dd \ln s} \label{schall},
  \end{equation}
where $s(z_H)=\frac{2\pi}{\kappa} \exp\{\frac32 A(z_H)\}$ stands for the entropy density. The quantities $T$, $c_s^2$ and $s$ refer to the boundary theory according to the holographic dictionary. The 
pressure is calculated via $p=\int \dd T s(T)$ with the side condition $p(T=0)=0$.

\subsection{Probe vector mesons} \label{sec2-2}
Additionally, we study the behavior of ``probe vector mesons'' which are not back-reacted, since they are solely meant to probe the background. We use the standard action in the Einstein frame 
(cf.~\cite{KKSS, braga, braga2, steph}; note the difference to the string frame action where an additional factor $e^{-\Phi}$ shows up)
  \begin{equation}
   S_V \propto \int \! \dd^5 x \sqrt{g} F^2, \label{mesonwirkung}
  \end{equation}
where $F^2$ is the squared field strength tensor of a $U(1)$ vector field $\mathscr{A}$. The equation of motion follows, after some manipulations \cite{col12}, as one-dimensional Schr\"odinger type 
equation \cite{ich2017}
  \begin{equation}
   \li(\Dp_{\xi}^2 -(U_T-m_n^2)\ri) \psi =0, \label{schr}
  \end{equation}
where $\Dp_{\xi}\equiv (1/f) \Dp_z$ and 
  \begin{equation}
   U_T=u_T f^2+\frac12 \mathcal{S}_Tff'
  \end{equation}
with the Schr\"odinger equivalent potential
  \begin{equation}
   u_T=\frac12 \mathcal{S}_T'+\frac14 \mathcal{S}_T^2, \quad \mathcal{S}_T \equiv \frac12 A'-\frac23 \Phi' \label{defs}.
  \end{equation}
In general, $A$ and $\Phi$ depend on both, $z$ and $z_H$. To distinguish the vacuum case ($T=0$: $A_0(z)$, $\Phi_0(z)$, $f(z)=1$) from the non-zero temperature case ($T>0$: $A(z,z_H)$, $\Phi(z,z_H)$ 
\blau{and $\Phi_H \equiv \Phi(z_H,z_H)$}, $f(z,z_H) \leq 1$) we add the label 0 to $A$ and $\Phi$. In case of zero temperature, $\xi=z$, the Schr\"odinger equivalent potential $U_0:=U_{T=0}$ is given 
by
  \begin{equation}
   U_0=u_0:=\frac12 \mathcal{S}_0'+\frac14 \mathcal{S}_0^2, \quad \mathcal{S}_0 \equiv \frac12 A_0'-\frac23 \Phi_0' \label{defs2}.
  \end{equation} 
Furthermore, $m_n$ denotes the masses of normalizable modes as solutions of (\ref{schr}) with $n=0,1, 2, \cdots$ as the quantum number of radial excitations. Of course, one has to be aware that 
the limit $U_T(z_H \to \un) \to U_0$ is continuous as stressed above. \\
The action (\ref{mesonwirkung}) is obviously flavor-blind, i.e.~it is not specific for light-quark or heavy-quark or light-heavy quark vector mesons. Instead, in order to describe different flavors, 
one must have different holographic backgrounds, e.g.~adjusted at $T=0$. From (\ref{schr}, \ref{defs}) one infers that only $U_0$, that is a special combination of $A_0(z)$ and $\Phi_0(z)$, is 
relevant. Thus, \cite{KKSS, braga, braga2, steph} tune different shapes and parameters of $U_0(z)$ accordingly to receive the wanted vector meson spectra -- partially also decay widths -- for 
$\rho$/$\omega$ mesons and charmonia and bottomonia at $T=0$ and employ them afterwards to meson melting phenomena at $T>0$.\\
In contrast, in our approach, the background is generated dynamically with emphasis on thermodynamics encoded in $p(T)$, $s(T)$, $c_s^2(T)$ etc.~at $T>0$. The wanted thermodynamics thereby can refer 
to various QCD scenarios with various flavor contents and/or chiral limit or heavy-quark limit as well. Therefore, it is a priori not clear to which of the flavor contents the action 
(\ref{mesonwirkung}) can be attributed or whether it is a purely fiducial test quantity. Nevertheless, despite the mentioned drawback of (\ref{mesonwirkung}), we are going to analyze whether and 
which normalizable solutions of (\ref{schr}) exist on backgrounds generated by a few-parameter dilaton-potential $V(\Phi)$.\footnote{A minimalistic way to include a scale in (\ref{mesonwirkung}), 
which may be linked to light or heavy flavors, would be to add a gauge symmetry breaking term $\propto M^2 \mathscr{A}^2$ \cite{braga2}. A much more refined improvement of (\ref{mesonwirkung}) is 
required to suitable flavor dependent quark masses and condensates.} \\
By employing the field equations, we find the following relation for $U_0(z)$:
  \begin{equation}
   U_0 = \frac{17}{48} {A_0'}^2 +\frac13 A_0'\Phi_0' +\frac13 e^{A_0}(\Dp_{\Phi} V-\frac16V). \label{u0}
  \end{equation}
In Section \ref{sec4-1} we study the probe vector meson spectrum over the background determined by solutions of (\ref{fgl1}-\ref{fgl4}) within $U_0$ from (\ref{u0}), while in Section \ref{sec4-2} an 
ansatz for $U_0(z)$, which facilitates a certain mass spectrum, is used as an input for (\ref{fgl1}-\ref{fgl4}) to figure out the related thermodynamics and phase structure.

\section{Thermodynamic scenarios} \label{sec3}
\subsection{Selected examples} \label{sec3-1}

\begin{figure}
      \includegraphics[scale=.37]{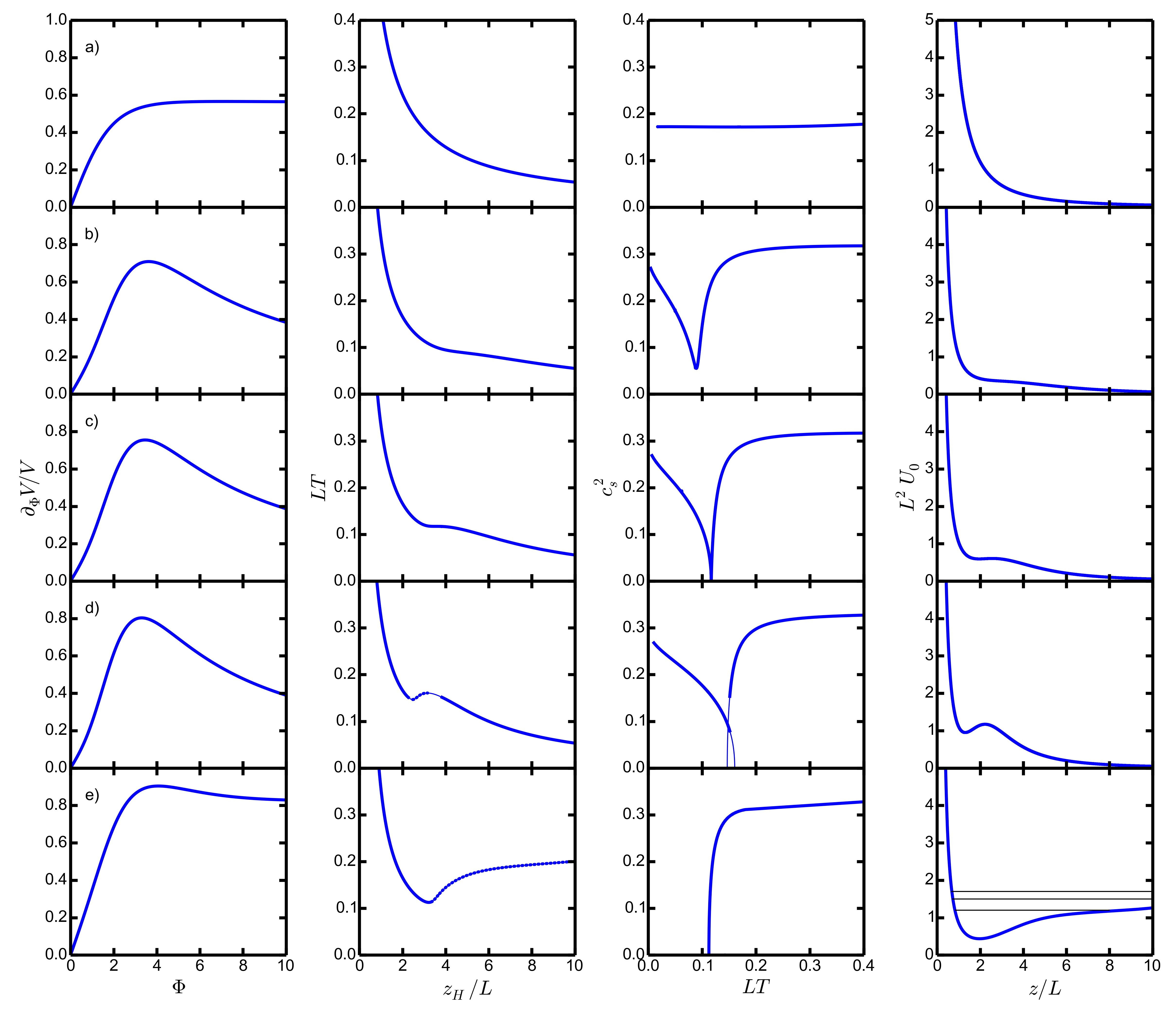}
      \caption{Selected examples of parameter choices (see Table \ref{tab1}) for the dilaton potential (\ref{kaninchen}) (left column, for $\Dp_{\Phi} V/V$), the resulting temperature $LT$ as a 
function of horizon position $z_H/L$ (second column), velocity of sound squared $c_s^2$ as a function of temperature $LT$ (third column) and the Schr\"odinger potential $L^2U_0$ as a function of 
$z/L$ (right column). The dilaton potential (\ref{kaninchen}) serves as an input to solve the field equations (\ref{fgl1}-\ref{fgl3}) and obtain $T(z_H)$ from (\ref{temp}), $c_s^2(T)$ from 
(\ref{schall}) combined with (\ref{temp}), and $U_0(z)$ from (\ref{u0}).} \label{abb1}
     \end{figure}

To illustrate the systematics of the thermodynamics related to the dilaton potential $V$ we choose the three-parameter ansatz\footnote{We follow \cite{mimi, fin1, pufu}. The relation to \cite{exp1, 
exp2, kir1, kir2, kir3} is discussed in \cite{boschi}. \blau{Two generalizations of (15) are considered in Appendix C.
Further parameterizations within that bottom-up approach have been}
\blau{ considered, e.g.\ in \cite{debye, hydro2}
for selected (fixed) coefficients. Our intention is to study the 
impact of the coefficients on the emergent phase structures. For that purpose, 
a three-dimensional parameter space is suitable for an easy illustration.
The 1-R charge black hole model is an example of a top-down approach,
which however is considered in \cite{critelli} as not suitable for direct
applications to relativistic heavy-ion collisions.}}
  \begin{equation}
   -L^2V(\Phi) = 12 \cosh (\gamma \Phi) +a\Phi^2 +b\Phi^4 \label{kaninchen}
  \end{equation}
because we can go through several thermodynamic scenarios (see Figure~\ref{abb1}) by changing smoothly the values of parameters $\gamma$, $a$ and $b$, which are related by $-L^2 m_{\Phi}^2 = 
12\gamma^2+2a$ to the dilaton mass parameter $m_{\Phi}^2$. The conformal dimension $\Delta$ is accordingly $\Delta=2+\sqrt{4-12\gamma^2-2a}$.

  \begin{figure}
      \cen{\includegraphics[scale=.37]{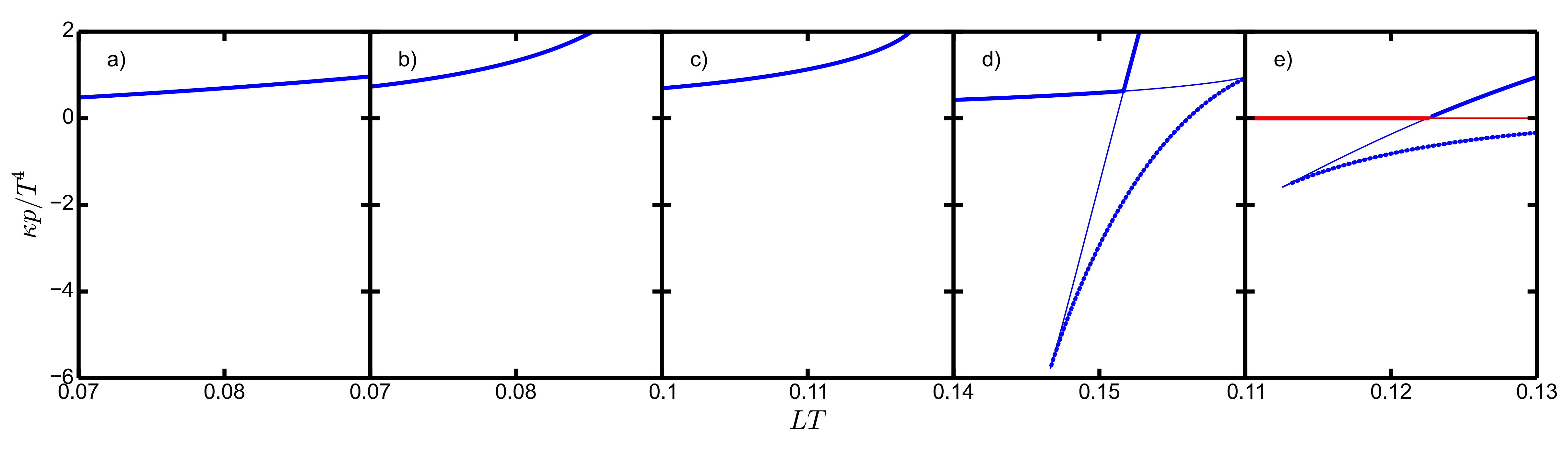}
      \caption{Scaled pressure $\kappa p/T^4$ as a function of $LT$ for the selected examples shown in Figure~\ref{abb1}. The red lines in panel e) are for the thermal gas solution ($f=1$), where $p 
= 0$ is employed (red lines). Meta (unstable) sections are depicted by thin (dotted) curves.} \label{abb2}}
     \end{figure}

  \begin{table}
      \cen{\begin{tabular}{c c c c c l}
       \hline
       Example & $\gamma$ & $a$ & $b$ & $\Delta$ & Transition \\
       \hline
       a) & 0.56 & -0.077 & 0 & 2.63 & none \\
       b) & 0 & 1.155 & 0.18 & 3.3 & cross-over \\
       c) & 0 & 1.155 & 0.20 & 3.3 & second-order \\
       d) & 0 & 1.155 & 0.25 & 3.3 & first-order \\
       e) & 0.83 & -2.69 & 0 & 3.06 & Hawking-Page \\
       \hline
      \end{tabular}}
      \caption{Parameter selection yielding the examples in Figure~\ref{abb1} and the characterization of the thermodynamic features.} \label{tab1}
     \end{table}

As already remarked in \cite{mimi}, quite featureless dilaton potentials $V(\Phi)$ can lead to fairly different thermodynamic features. Since the field equations 
(\ref{fgl1}-\ref{fgl3}) can be rearranged to display only a sensitivity to $\Dp_{\Phi} V/V$ as a function of $\Phi$, we plot this key quantity in the left column of Figure~\ref{abb1}. Example a) does 
not exhibit any features: $T(z_H)$ is monotonously decreasing, $c_s^2(T)$ is increasing, and $U_0$ has no minimum, meaning that normalizable modes as probe vector mesons do not exist at all.
Examples b) - d) exhibit $\Dp_{\Phi}V/V$ with a pronounced maximum which becomes gradually higher. In case b), $T(z_H)$ is monotonously dropping, albeit with a shallow near-flat section; it causes a 
pronounced minimum of the sound velocity squared; $U_0$ does not allow for any normalizable states due to the lacking minimum. That case is classified as cross-over.
Case c) features a lifted maximum of $\Dp_{\Phi}V/V$, resulting in a flat section of $T(z_H)$, and the sound velocity drops at a certain temperature to zero, thus representing an example of a 
second-order phase transition. The Schr\"odinger equivalent potential $U_0(z)$ has here a very shallow minimum, i.e.~probe vector mesons as normalizable modes cannot show up. 
Lifting the maximum of $\Dp_{\Phi} V/V$ further (case d)), $T(z_H)$ shows a local minimum that is connected to an inflection point, implying metastable states, unstable states and spinodales as well. 
That becomes most evident by the sound velocity squared $c_s^2(T)$ and the pressure $p(T)$ (see Figure~\ref{abb2}): metastable states are depicted by thin solid curve sections and the unstable ones 
by dotted sections. Clearly, states with $c_s^2<0$ cannot be realized in nature. All these features classify a first-order phase transition, for which $U_0(z)$ exhibits a local minimum; it is - for 
the given parameters - still too shallow to accommodate vector mesons.
Only if $\Dp_{\Phi}V/V$ exceeds $\sqrt{2/3}$ (case e)), the global minimum of $U_0(z)$ allows for meson states, depicted by horizontal lines. At the same time, $T(z_H)$ has a global minimum pointing 
to a Hawking-Page (HP) phase transition. Let be $T_{\min}=\min T$ at $z_H^{\min}$. Then, the branch for $z_H <z_H^{\min}$ is stable (for $p>0$) and metastable (for $p<0$), while the branch for 
$z_H>z_H^{\min}$ is unstable, and its free energy is above the thermal gas solution (see \cite{kir1} for the related construction) which applies for $0<T<T_{\min} <T_{\rm c}<\un$, where $T_{\rm c}$ 
(slightly above $T_{\min}$) is the first-order phase transition temperature. (While $T_{\min}$ follows as minimizer of (\ref{temp}) over $z_H$, $T_{\rm c}$ is obtained by the loop construction 
sketched in the rightmost graphs of Figure~\ref{abb2}.) The velocity of sound drops to zero at $T_{\rm c}$. \\
These examples are selected to have a survey on the thermodynamics and holographic quantities which can be uncovered by (\ref{kaninchen}). This information is refined in Subsection~\ref{sec4-1}, 
where we provide a systematic scan through the planes $\gamma=$constant and $b=$constant. \\
Figures \ref{glas} and \ref{3d} exhibit that region in the $\gamma$-$a$-$b$ parameter space, where a first-order or HP phase transition occurs; and in Figure~\ref{3d}, curves on which a second-order 
phase transition happens are depicted too. Thanks to the three-parameter potential (\ref{kaninchen}), the visualization of the parameter space structure is quite straightforward, while multi-parameter 
ans\"atze may lead to intricate structures. In the present case, the HP transition happens for $\gamma \geq 0.8$ for all BF permitted values of $a$ and $b$ as well, as exhibited in the left panel 
of Figure~\ref{3d}. For $\gamma <0.8$, the blue areas depict the first-order phase transition, which are limited by the second-order transition (red curves). Further left (gray parts of the panels 
with constant $\gamma$ or $a$ or $b$ cross sections), a cross-over occurs, which turns smoothly into a featureless behavior for smaller values of $b$.

\subsection{Shaping the dilaton potential} \label{sec3-2}

For $\gamma=0.568$, $a=-1.92$ and $b=-0.04$, the potential (\ref{kaninchen}) reproduces the Lattice QCD data \cite{baza} of $c_s^2(T)$ fairly well when adopting the scale setting parameter 
$L^{-1}=1990$~MeV. These data and our fit are restricted to the range 125~MeV$<T<$450~MeV. The dependence $LT$ vs.~$z_H/L$ looks nearly the same as in Figure~\ref{abb1}, case b), while the 
combination 
$\Dp_{\Phi}V/V$ as a function of $\Phi$ exhibits a local maximum at $\Phi \approx 1.4$ and a local minimum at $\Phi \approx 5$ and stays below 0.6, i.e.~these details are somewhat different from that 
displayed in Figure~\ref{abb1}, case b), left panel. The successful description of the sound velocity squared can be taken as argument for considering the potential (\ref{kaninchen}) useful for 
catching essential QCD features. \\
We refrain from further quantitative comparison with QCD thermodynamics and refer the interested reader to \cite{kir1, mimi, debye, yaresko, yaresko-knaute}, for example, where the thermodynamics of 
Yang-Mills and 2+1 flavor QCD with physical quark masses is considered. The essence is to employ multi-parameter ans\"atze of the dilaton potential with the aim to reproduce the Lattice QCD 
thermodynamics results as good as possible.\footnote{For a more generic study of the potential and the related RG flow, cf.~\cite{kir2017}.} \pagebreak
Information of QCD is thus imported and mapped 
in a 
cumulative manner on $V(\Phi)$ within such a bottom-up approach without explicit reference to quarks and gluons and their masses, colors, flavors, couplings etc.\footnote{In contrast, the IHQCD model 
\cite{exp1, exp2, kir1, kir2, kir3} aims at anchoring fundamental QCD features in the chosen ansatz from the beginning; the approaches in \cite{debye, topdown}, following the 1-R charge black hole 
(1RCBH) model \cite{rmodel1, rmodel2, rmodel3, rmodel4, rmodel5}, in turn are string theory driven.} Having successfully accomplished the shaping of $V(\Phi)$, one can proceed to derive further 
quantities, such as viscosities \cite{pufu, rocha}, diffusion constants \cite{rouge, rouge2} etc.~as predictions. By extending the model (\ref{wirkung}) by further fields, e.g.~a Maxwell type $U(1)$ 
gauge field \cite{wolfe, yaresko-knaute, knaute}, one may address non-zero chemical potential effects to access the phase diagram and issues of a critical point. Here, susceptibilities serve as 
crucial further information to be imported from QCD too. \\
However, our present goal here is to answer the question whether one can describe -- within the modeling (\ref{wirkung}, \ref{mesonwirkung}) -- at the same time the LHC relevant QCD thermodynamic 
features, i.e.~a cross-over at about 155~MeV \cite{borsa, baza}, and a proper in-medium behavior of hadrons, i.e.~probe vector mesons as representatives thereof. By a proper 
``in-medium behavior'' we mean that at chemical freeze-out temperature of about 155~MeV \cite{alice}, hadrons do exist with hardly medium-modified properties. Otherwise, the famous thermo-statistical 
interpretation of hadron multiplicities in ultra-relativistic heavy-ion collisions \cite{alice} would be invalidated. 

  {\begin{figure}
   \cen{\includegraphics[scale=.4]{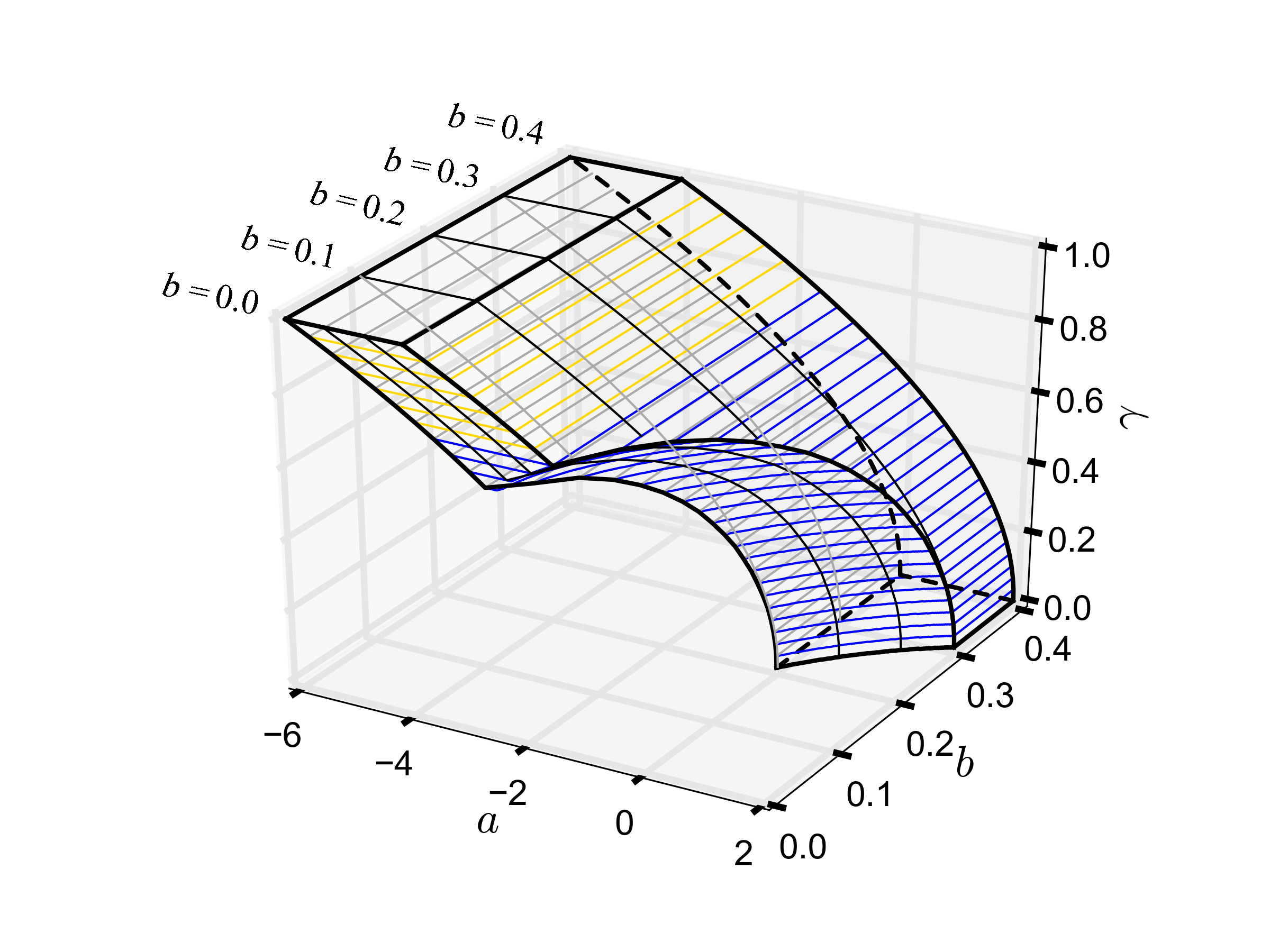}
   \caption{Plot of the region in the $\gamma$-$a$-$b$ parameter space, where a HP or first-order phase transition in the BF allowed range occurs for the dilaton potential (\ref{kaninchen}).  
Cross sections of constant values of $\gamma$, $a$ and $b$ are exhibited in Figure~\ref{3d}.} \label{glas}}
  \end{figure}

  \begin{figure}
   \cen{\includegraphics[scale=.36]{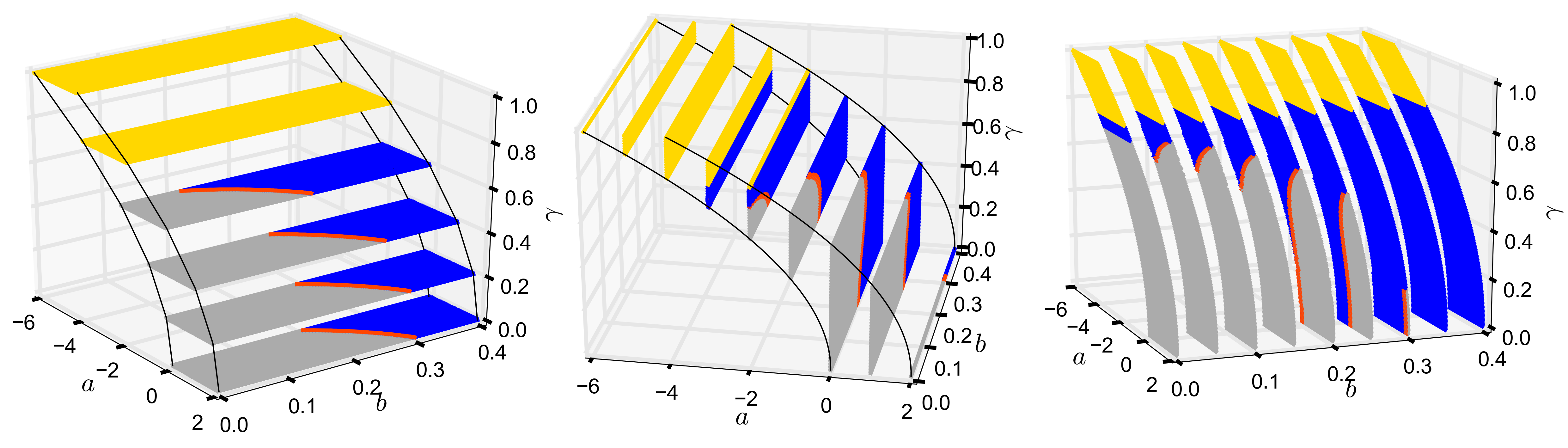} 
   \caption{Plot of parameter ranges of the dilaton potential (\ref{kaninchen}) for first-order phase transition (blue areas) or HP transition (yellow areas) for a series of selected constant values 
of $\gamma= 0, \cdots$, 1 (left), $a=-6, \cdots, 2$ (middle) and $b=0, \cdots, 0.4$ (right). The red curves mark the second-order phase transition ranges; beyond, the transition turns into a 
cross-over, followed by a featureless behavior (dark gray areas). The BF bound restricts the values of $a$ to the strip $-6 \gamma^2 < a <2 -6 \gamma^2$ (depicted by the two black solid curves in 
left and middle panels). The panels continue to larger values of $b$ and $\gamma$ without changes.} \label{3d}}
  \end{figure}}
  
The answer to the posed problem seems to be negative. Hints come, for example, from \cite{col12, col09, gherghetta, kapusta, bk}, where the melting (disappearance) of hadrons was found to happen at 
temperatures significantly below 155~MeV. While \cite{ich2017} offers an avenue to remedy such an insanity, the given framework of (\ref{wirkung}, \ref{mesonwirkung}) seems to be too restricted and 
calls for extensions. Leaving the latter ones for separate work, we try to find a loophole to join the cross-over thermodynamics and suitable probe vector meson states. Prior to that, however, we 
attempt to clarify the systematics of thermodynamic features in the spirit of the last column of Tab.~\ref{tab1} in relation to the dilaton potential.

\subsection{Beyond the adiabatic approximation} \label{sec3-3}

\begin{figure}
      \cen{\includegraphics[scale=.44]{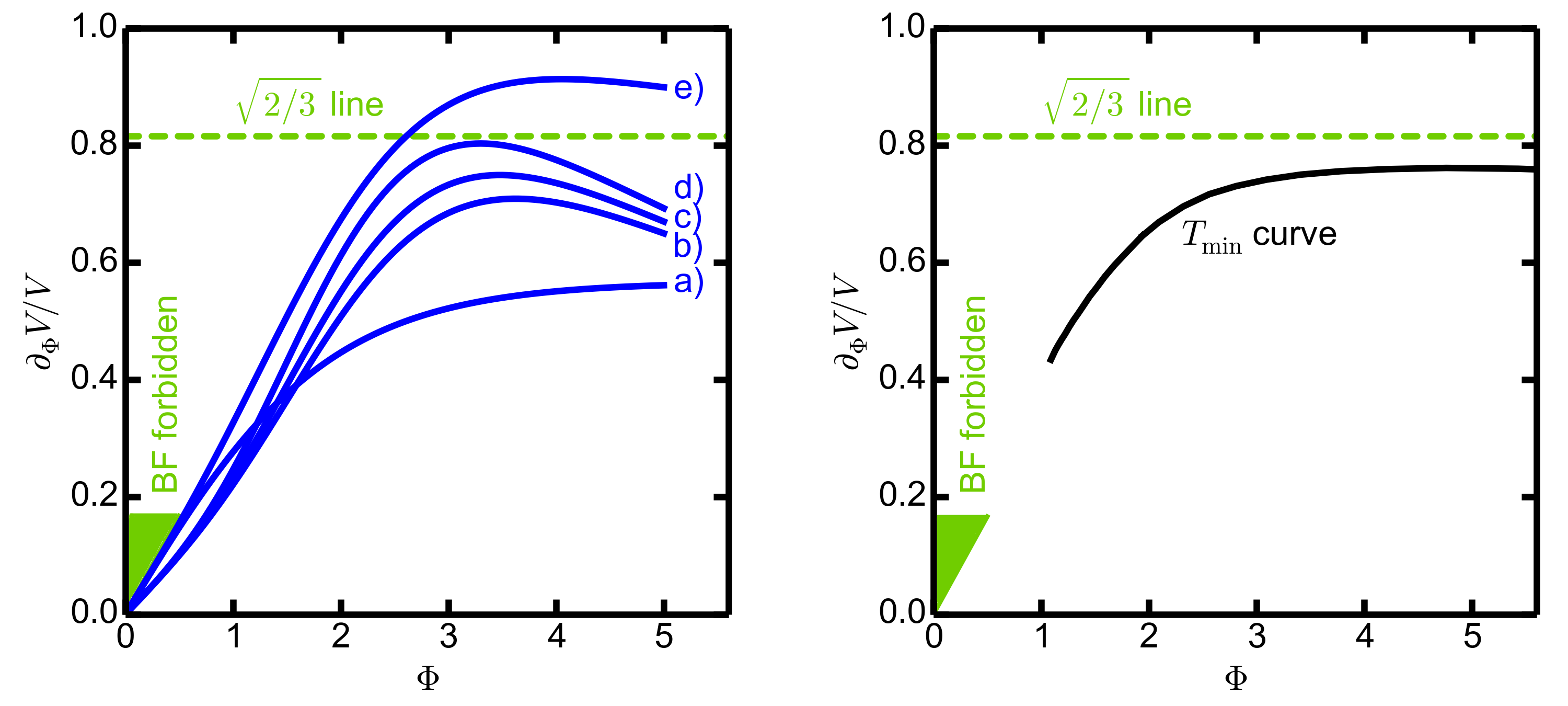}
      \caption{Left: Comparison of the dilaton potentials of examples a) - e) in Figure~1 for $\Phi=0\cdots5$. The horizontal line is the adiabatic criterion (16).
Right: \blau{Comparison of Gubser's criterion and the $T_{\min}$ curve which shows the minimum value of $\partial_{\Phi} V/V$ as a function of $\Phi_H$ for which $T(z_H)$ displays a minimum pointing 
to a first-order or HP transition. If $\partial_{\Phi} V/V$ as a function of $\Phi$ exceeds the $T_{\min}$ curve, then $T(\Phi_H)$ exhibits a minimum; otherwise, if $\partial_{\Phi} V/V$ stays below 
the $T_{\min}$ curve, $T(\Phi_H)$ decreases monotonously which points to a cross-over transition or to featureless thermodynamics. (In detail, for the dilaton potential (15) with 
$\gamma=b=0$ and a given value of $a$ with $0 \leq a\leq 2$, $T(\Phi_H)$ is computed as well as the position of its minimum $\Phi_H^{\min}(a)$. The $T_{\min}$ curve then is the connection of all 
points $\left(\Phi_H^{\min}, \partial_{\Phi} V/V(\Phi_H^{\min})\right)$ for running values of $a$.)}} \label{abb3}}
     \end{figure}

The authors of \cite{mimi} derived the relation (henceforth called Gubser's adiabatic criterion)
  \begin{equation}
   c_s^2 \approx \frac13-\frac12 \li(\frac{\Dp_{\Phi}V}V \ri)^2, \label{X1}
  \end{equation}
where `$\approx$` indicates the validity in adiabatic approximation.\footnote{One may exploit (\ref{X1}) to get a suitable form of $V(\Phi)$ by adjusting a parametrized 
ansatz for $h(T)=\frac13-c_s^2$. As an example we mention $h(T)=(T/T_1)^n/[1+(T/T_2)^{n+2}]$ with optimum parameters $(T_1/$MeV, $T_2/$MeV, $L^{-1}/$MeV, $n)=(141.1, 126.7, 1799, 11.17)$ for the data 
\cite{baza} and $(139.5, 115.5, 1714, 5.5)$ for \cite{borsa}. $V(\Phi)$ is then determined by Equation~(\ref{T1}) assuming ${\Dp A}/{\Dp z_H}={\Dp \Phi}/{\Dp z_H}=0$. The combination $\Dp_{\Phi}V/V$ 
as a function of $\Phi$ exhibits then a maximum of about 0.6 at $\Phi \approx 3$ and declines towards zero at $\Phi \approx 5$. The according shape of $V(\Phi)$ can be obtained by (\ref{kaninchen}) 
for $\gamma=0$, $a=2$ and $b=-0.03$. Clearly, this finding corroborates our above statement at the beginning of Subsection~\ref{sec3-2}. Otherwise, one sees that quite different shapes of 
$V(\Phi)$ are suitable for reproducing the Lattice QCD data within the given uncertainty range and, in particular, within the restricted temperature interval. Forthcoming precision data are needed to 
constrain better the dilaton potential in such a bottom-up approach.}
The formula implies that for $\Dp_{\Phi}V/V > \sqrt{2/3}$ the sound velocity becomes imaginary, thus pointing to a first-order 
phase transition, either as a standard construction $\rm \grave{a}$ la example d) or the HP transition $\rm \grave{a}$ la example e) in Figure~\ref{abb2}. To systematize the various thermodynamic 
scenarios, we plot $\Dp_{\Phi} V/V$ of the above examples a) - e) in one diagram, see Figure~\ref{abb3}. Based on such a comparison, the impact of $\Dp_{\Phi}V/V$ on the thermodynamics can be 
summarized qualitatively as follows:
If $\Dp_{\Phi}V/V$ reaches the value $\sqrt{2/3}$ (or somewhat below, depending on the concrete $V(\Phi)$), $T(z_H)$ forms a local extremum, where the nearest 
one to the boundary becomes a minimum. So if $\Dp_{\Phi}V/V$ intersects the $\sqrt{2/3}$ line once, we have a global minimum of $T(z_H)$ and a HP phase transition. Otherwise, if $\Dp_{\Phi}V/V$ 
intersects twice, $T(z_H)$ forms a local minimum followed by a local maximum and we have a first-order phase transition. A second-order phase transition arises if $\Dp_{\Phi} V/V$ touches the 
$\sqrt{2/3}$ line. \pagebreak
Additionally, each extreme point of $\Dp_{\Phi}V/V$ implies an inflection point of $T(z_H)$, i.e.~a cross-over is generated by a maximum of $\Dp_{\Phi}V/V$ whose altitude stays below $\sqrt{2/3}$. \\
To formalize these findings we derive in Appendix \ref{anhang_A} the relation
  \begin{eqnarray}
   \frac1T \frac{\dd T}{\dd z_H} &=& \frac12 \frac{V}{\Dp_{\Phi} V} \li(\li(\frac{\Dp_{\Phi} V}V \ri)^2-\frac23 \ri) \Phi' \nonumber \\ && +\frac{\Dp A}{\Dp z_H} + \li(\frac32 \frac{\Dp A'}{\Dp z_H} 
+\Phi' \frac{\Dp \Phi}{\Dp z_H} \ri)\frac{\Dp_{\Phi} V}V, \label{T1}
  \end{eqnarray}
where $\Phi' \equiv (\Dp_z \Phi(z,z_H)) \mid _{z=z_H}$, $A' \equiv (\Dp_z A(z,z_H)) \mid _{z=z_H}$.

\begin{figure}
   \cen{\includegraphics[scale=.55]{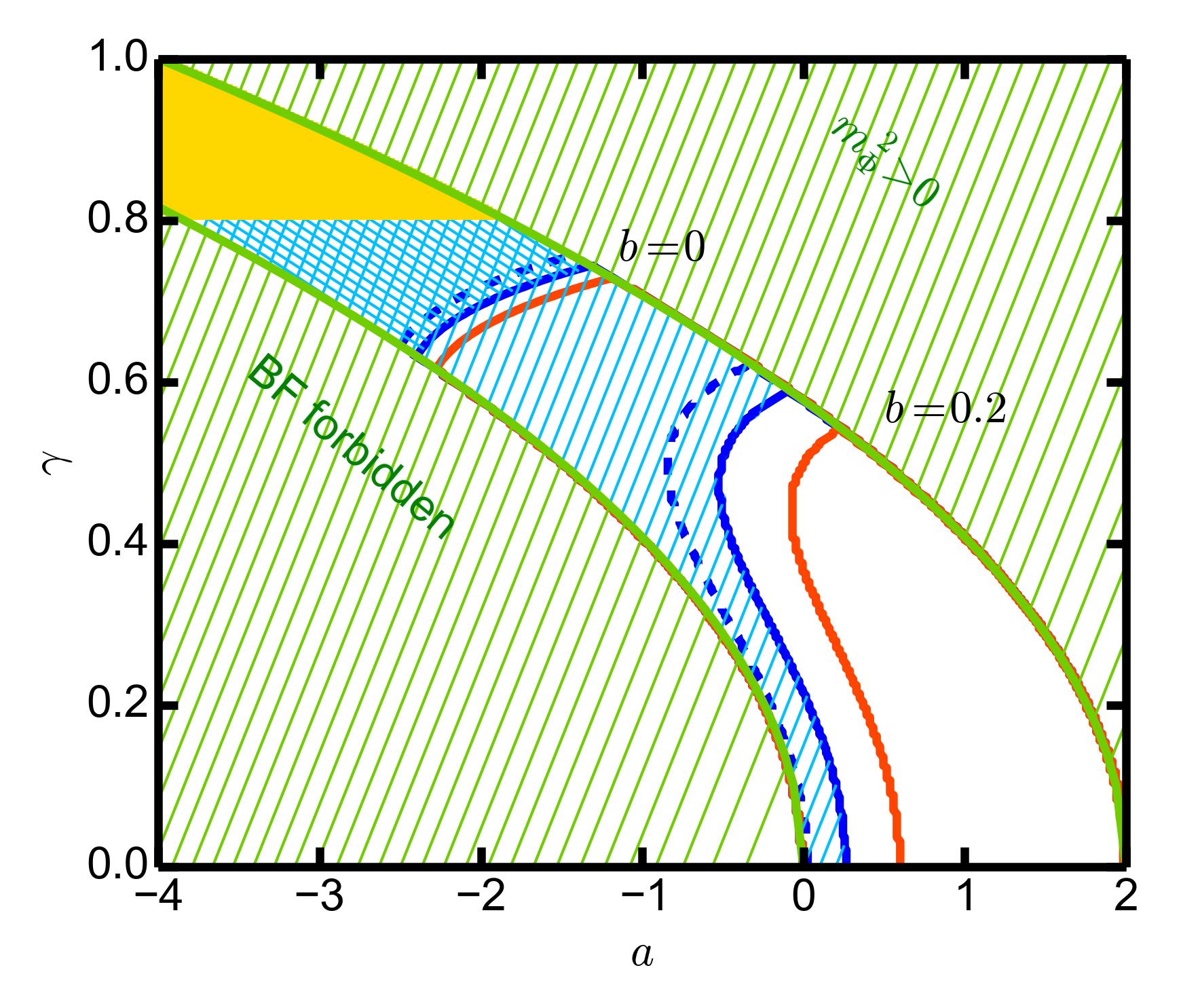}
   \caption{Projection of border lines (for $b=0$ and $b=0.2$) of parameter regions on $\gamma$ vs.~$a$ plane, where $T(z_H)$ has a minimum (blue hatched regions, left and above to solid blue curves), $U_0$ has a minimum (left and above to red curves), and $\Dp_{\Phi} V/V$ is greater than $\sqrt{2/3}$ (left and above to dashed blue curves) for the dilaton potential (\ref{kaninchen}); if $b$ and therefore the slope of $\Dp_{\Phi} V/V$ increase, the difference between Gubser's adiabatic criterion (blue dashed) and the $T_{\min}$ curve (solid blue) becomes larger. In the blue double-hatched region, $\gamma$ is great enough to have a first-order phase transition for all $a$ and $b$. The region of a HP transition is in yellow. The green hatched regions are excluded due to BF.} \label{abb3a}}
  \end{figure}

Given the facts that (i) $T(z_H \to 0) \to \frac1{\pi z_H}$ \cite{cherman, hohler}, (ii) the monotonous behavior of $\Phi(z,z_H)$ as a function of $z$ with $\Phi'>0$, and (iii) the above quoted 
asymptotic behavior of $V(\Phi)$ at small $\Phi$ (implying $\Dp_{\Phi} V/V = m_{\Phi}^2 L^2\Phi/12$), one recognizes from the first line of (\ref{T1}) that the slope $\dd T/\dd z_H$, which is negative 
at small $z_H$, can turn into a positive one, once $\Dp_{\Phi}V/V >\sqrt{2/3}$ is reached, indicating a local or global minimum of $T(z_H)$. Understanding ``adiabatic approximation'' as a situation 
where $\Dp A/\Dp z_H$, $\Dp A'/\Dp z_H$ and $\Dp \Phi/\Dp z_H$ are small, one thus recovers Gubser's adiabatic criterion. Otherwise, the second line of (\ref{T1}) provides corrections. In fact, in 
example d), $\Dp_{\Phi} V/V$ stays below the $\sqrt{2/3}$ line but facilitates a first-order phase transition.
(The relation of $T(z_H)$ to transitions is discussed in \cite{exp1, exp2}: in essence, a minimum of $T(z_H)$ points to a first-order phase transition, since $s(z_H)$ is a monotonously increasing 
function.) The corrections give eventually a border line (called $T_{\min}$ curve) for each type of dilaton potential which is determined by calculating the minimum value of $\Dp_{\Phi} V/V$ (as a 
function of $\Phi$ and depending on all parameters denoted shortly by $\vec p \, $) such that $T (z_H)$ forms a minimum.
Systematic numerical analyses with the dilaton potential (\ref{kaninchen}) show that this line is shifted down if $\Dp_{\Phi} V/V$ as a function of $\Phi$ becomes steeper when varying the parameters 
$\vec p$ in $V(\Phi; \vec p)$. 
The right part of Figure \ref{abb3} shows an example of such a line and the dependence of the difference between Gubser's criterion and the $T_{\min}$ curve as a function of $\Phi$. This is further 
visualized in Figure~\ref{abb3a} for two special parameters, $b=0$ and 2, in the projections on the $\gamma$-$a$ plane: the offset of the regions determined by (\ref{X1}) and the true onset of a 
first-order phase transition increases with $b$; in addition, the region where the Schr\"odinger equivalent potential $U_0$ displays a minimum is shown by the red curves.

\section{Schr\"odinger potential} \label{sec4}
\subsection{Adiabatic approximation to Schr\"odinger equivalent potential} \label{sec4-1}

\begin{figure}
   \includegraphics[scale=.48]{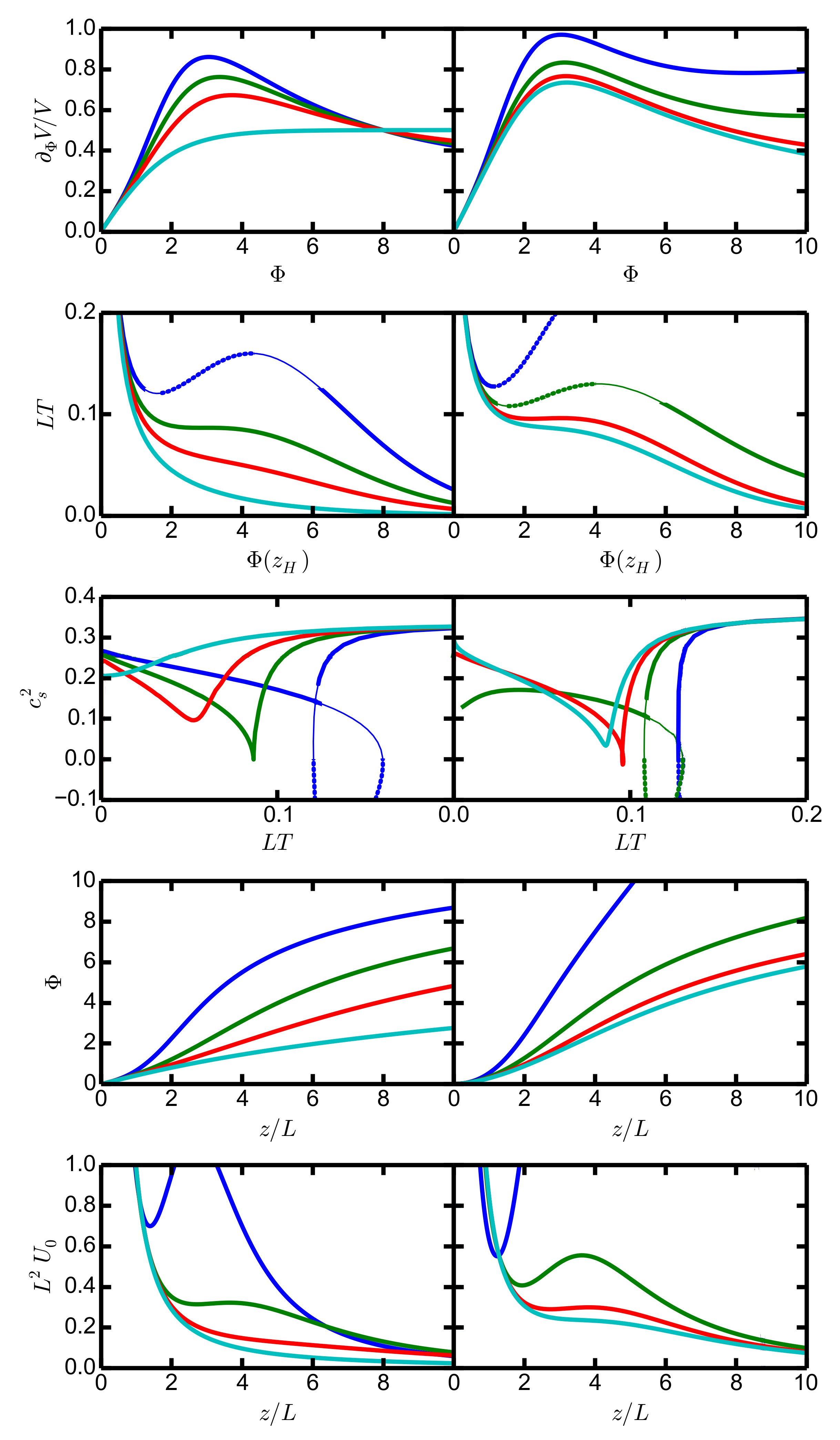}
   \caption{A scan through the planes $\gamma=0.5$, $a=0$ (left column, compare Figure~\ref{3d}-left) and $b=0.2$, $\Delta=2$ (right column, compare Figure~\ref{3d}-right) by selecting a few values 
of the remaining parameters of (\ref{kaninchen}) to exhibit the induced change of $\Dp_{\Phi}V/V$ as function of $\Phi$, $LT$ as a function of $\Phi(z_H)$, $c_s^2$ as a function of $LT$, $\Phi$ as a 
function of $z/L$ and $L^2U_0$ as a function of $z/L$ (from top to bottom). The line style is as in Figure~\ref{abb1}; values of the remaining parameters:}
  \label{laufen}
    \begin{tabular}{l c @{\,\,\,\,} c @{\,\,\,\,} c @{\,\,\,\,} c}
    \hline
    line color & blue & green & red & cyan \\
    \hline
    $b$ (left column) & 0.3 & 0.18 & 0.1 & 0 \\
    $a$ (right column) & 0.83 & 0.65 & 0.5 & 0.3 \\
    \hline
   \end{tabular}
  \end{figure}

The relationship between the situation of $T=0$ and features at $T>0$ has been stressed in \cite{exp1, exp2}. Here, we envisage a relation of $U_0(z)$ and $T(z_H)$ in adiabatic approximation. In 
Appendix \ref{anhang_B} we derive the relation
  \begin{equation}
   9U_0 \cong \frac{17}{12}\frac{T'^2}{T^2} +\frac{28}3\pi T'+\frac{44}3\pi^2 T^2+ 2\li(2\pi T-\frac{T'}T\ri) \sqrt{\frac83 \li(\frac{T'^2}{T^2}+5\pi T'+4\pi^2T^2\ri)}. \label{final}
  \end{equation}
It is valid if, in a decomposition $A(z,z_H)=A_0(z)+a(z,z_H)$ and $\Phi(z,z_H)=\Phi_0(z)+\varphi(z,z_H)$, where $A_0$ and $\Phi_0$ denote the solutions of (\ref{fgl1})-(\ref{fgl3}) in case of 
$f=1$ and the terms $a(z,z_H)$ and $\varphi(z,z_H)$ are sub-leading and can be neglected. The Chamblin-Reall solution \cite{CR} with $V(\Phi) \propto \exp\{\gamma \Phi\}$ is an example, where 
$a=\varphi=0$ can be chosen, despite of $f(z,z_H) \leq 1$. \\
Equation (\ref{final}) is exact for the AdS-BH metric, i.e.~$A=-2\ln(z/L)$, $\Phi(z)=0$ and $f(z,z_H)=1-(z/z_H)^4$ which is generated by $V(\Phi)=-12/L^2$. Moreover, the left side 
converges against the right side, if (i) $z\to0$ because $A(z\to0,z_H) \to -2\ln(z/L)$ and $\Phi(z\to0,z_H) \to0$ for all $z_H$ and (ii) $z \to \un$ due to $A\to A_0$, $\Phi \to \Phi_0$ and the 
implied large values of $z_H$. \\
If we assume that $T(z_H)$ has a minimum at $z_H=z_H^{\min}$, (\ref{final}) yields $U_0'(z=z_H^{\min}) = \frac\pi3(43-8\sqrt6) T''>0$ meaning $U_0$ increases at the minimum position of 
$T$. Due to the AdS asymptotics of the Schr\"odinger potential, $U_0'(z \to 0) \propto - z^{-3}<0$, there has to be a minimum of $U_0(z)$ as well, at a position nearer to the 
boundary, i.e.~in the interval $0<z<z_H^{\min}$. While derived within the above approximations, and thus not as rigorous as a no-go-theorem, one could argue that a minimum of $T(z_H)$ (which 
is related to a HP or first/second-order transition) is consistent with a minimum of $U_0(z)$. The reversed clue (though not necessarily true in any case, cf.~Figure~\ref{abb1}c) is demonstrated 
by an example in the next subsection. \\
Before requiring a minimum of $U_0(z)$, let us consider the reason for the disappearance of the $U_0$ minimum for certain parameter settings. The UV region of $U_0(z)$ is supposed to be determined by 
the near boundary behavior of $A_0(z)$, while the IR behavior is supposed to be determined essentially by the dilaton field. If true, then a piecewise shape $\Phi_0(z) \propto z^{p+1}$ generates a 
contribution $\propto z^{2p}$ to leading order $U_0$ in the IR. If such a term is dominating, then $U_0 \propto z^{2p}$, i.e.~$p>0$ is needed arrive at a shape of $U_0$ in the IR with $\Dp U_0 / \Dp 
z>0$. To quantify such a rough consideration (which ignores the coupling of $\Phi$ and $A$ via (\ref{fgl2})) we have scanned through the parameter space of (\ref{kaninchen}) on two representative 
directions, see Figure~\ref{laufen}. In doing so, we see in fact that only in the region of a first-order or second-order or Hawking-Page phase transition (cf.~Figure~\ref{3d}) the rise of the 
dilaton field in $z$ direction is strong enough to enforce also the IR rise of $U_0$, meaning that only in such cases $U_0$ can exhibit a pronounced minimum which is the prerequisite to allow for 
normalizable solutions of (\ref{schr}) at $T=0$. Figure~\ref{laufen} offers a better understanding of the deformations of the various quantities, e.g.~$c_s^2(T)$ or $T(\Phi(z_H))$ or $U_0(z)$, under 
continuous changes of the dilaton potential parameters, thus supplementing Figure~\ref{abb1}.

\subsection{Requiring a minimum of $\boldsymbol{U_0(z)}$} \label{sec4-2}
\begin{figure}
      \cen{\includegraphics[scale=.4]{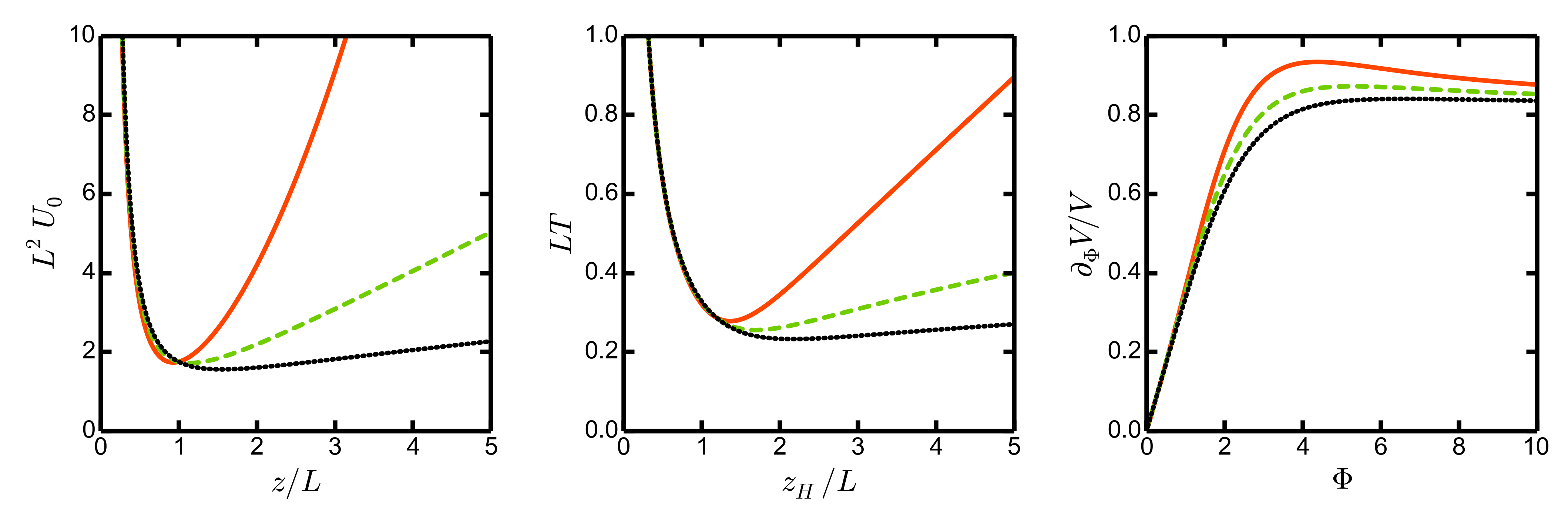}
      \caption{$L^2U_0$ as a function of $z/L$ (left panel), $LT$ as a function of $z_H/L$ (middle panel), and $\Dp_{\Phi} V/V$ as a function of $\Phi$ (right panel) for the input (\ref{u0p}). Red 
solid/green dashed/black dotted curves are for $p=2$/1/0.5.} \label{abb4}}
     \end{figure}

The above examples demonstrate that for many parameter choices of the dilaton potential $V(\Phi)$ the Schr\"odinger potential $U_0(z)$ does not exhibit a minimum and thus does not allow for modes 
which can be interpreted as probe vector mesons. Instead of deriving the background (warp factor and dilaton profile at $T=0$) from given dilaton potential, we start now with an ansatz for $U_0(z)$ 
such as to have a minimum. Assuming the latter one is sufficiently deep, normalizable modes would then be expected. Our ansatz for demonstrative purposes is
  \begin{equation}
   U_0(z) =\frac3{4z^2}+\li(\frac{z}L\ri)^p \frac1{L^2}, \label{u0p}
  \end{equation}
where the first term comes from the asymptotic warp factor at $z \to 0$; the second term facilitates the required minimum at $z_{\min}/L=(3/2p)^{\frac1{p+2}}$ 
with $L^2U_0^{\min}=(3/2p)^{\frac{p}{p+2}}(1+p/2)$. Equation (\ref{defs}) is solved at $f=1$ by
   \begin{equation}
   \mathcal{S}_0= 2 \frac{\dd}{\dd \hat z} \ln \li(c_1 \hat z^{-\frac12} \, _0F_1 \li(\frac{p}{p+2},\frac{\hat z^{p+2}}{(p+2)^2} \ri) + c_2 \hat z^{\frac32} \, _0F_1 \li(\frac{p+4}{p+2},\frac{\hat 
z^{p+2}}{(p+2)^2} \ri)\ri)
  \end{equation}
with $\hat z \equiv z/L$ and $c_1=1$ (due to AdS behavior at boundary $z \to 0$) and $c_2 =-\frac12 (p+2)^{\frac{p-2}{p+2}} \Gamma(\frac{p}{2+p})/ \Gamma(\frac2{p+2})$ (due to the assumption that 
(\ref{u0p}) is globally valid). The field equations (\ref{fgl1}-\ref{fgl3}) must be solved numerically to get $A(z)$, $\Phi(z)$ and $V(\Phi)$. The same $V(\Phi)$, which is supposed to be independent 
of $T$, is then used to derive $U_T(z, z_H)$ and $T(z_H)$. Figure \ref{abb4} exhibits such solutions for $p=0.5$, 1 and 2, where the latter value reproduces the soft-wall model \cite{KKSS} with a 
strictly linear Regge type spectrum $L^2m_n^2 = 4(n+1)$ for $n=0,1,2 \cdots$. 
The left panel is for $U_0(z)$ according to (\ref{u0p}), while the middle panel shows $T(z_H)$; the right panel displays $\Dp_{\Phi} V/V$ as a function of $\Phi$. There is a striking similarity of 
the curves $T(z_H)$ and $U_0(z)$ (isotopy referring to the monotony behavior) which we interpret as follows: a minimum of $U_0(z)$ is related to a minimum of $T(z_H)$, i.e.~a first-order phase 
transition  - here a HP transition, since it is a global minimum. The behavior of $\Dp_{\Phi} V/V$ as a function of $\Phi$ is in agreement with our assessments in Sections \ref{sec3-1} and 
\ref{sec3-3}. 

     \begin{figure}
      \cen{\includegraphics[scale=.4]{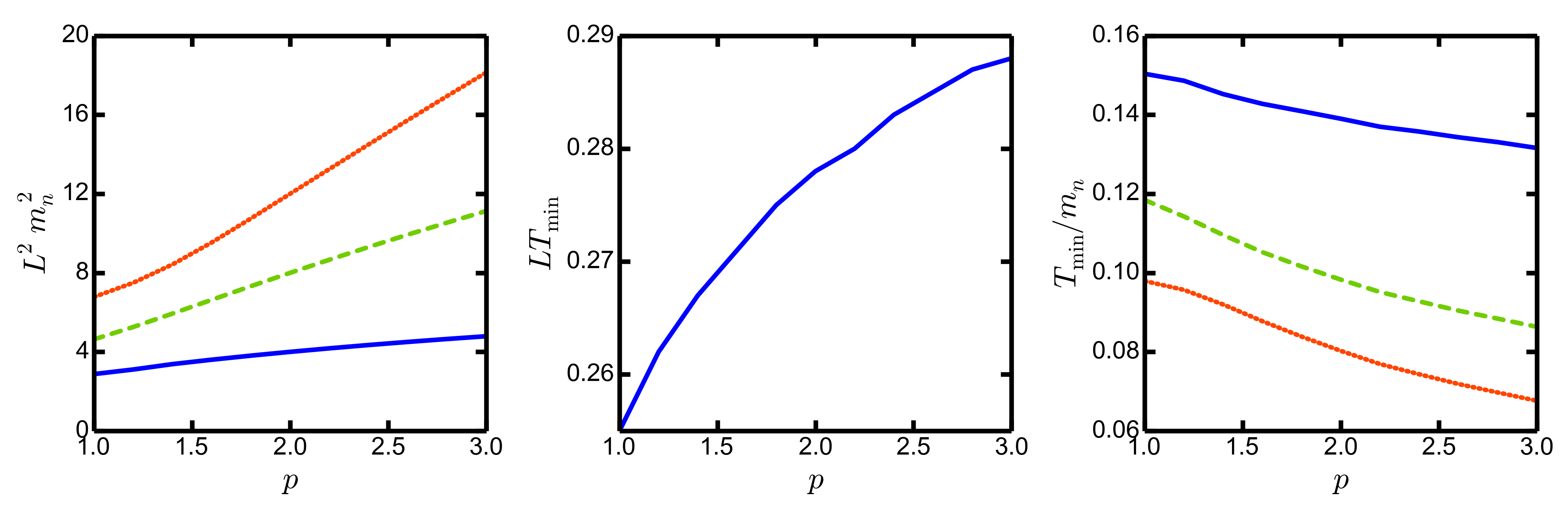}
      \caption{Scaled vector meson masses squared $L^2m_n^2$ as a function of $p$ (left panel), scaled minimum temperature $LT_{\min}$ as a function of $p$ (middle panel), and ratio $T_{\min}/m_n$ as 
a function of $p$ for the ansatz (\ref{u0p}). Blue solid/green dashed/red dotted for ground state/first excitation/second radial excitation (there is an infinite tower of excitations for all $p$).} 
\label{abb5}}
     \end{figure}

\begin{figure}
      \cen{\includegraphics[scale=.4]{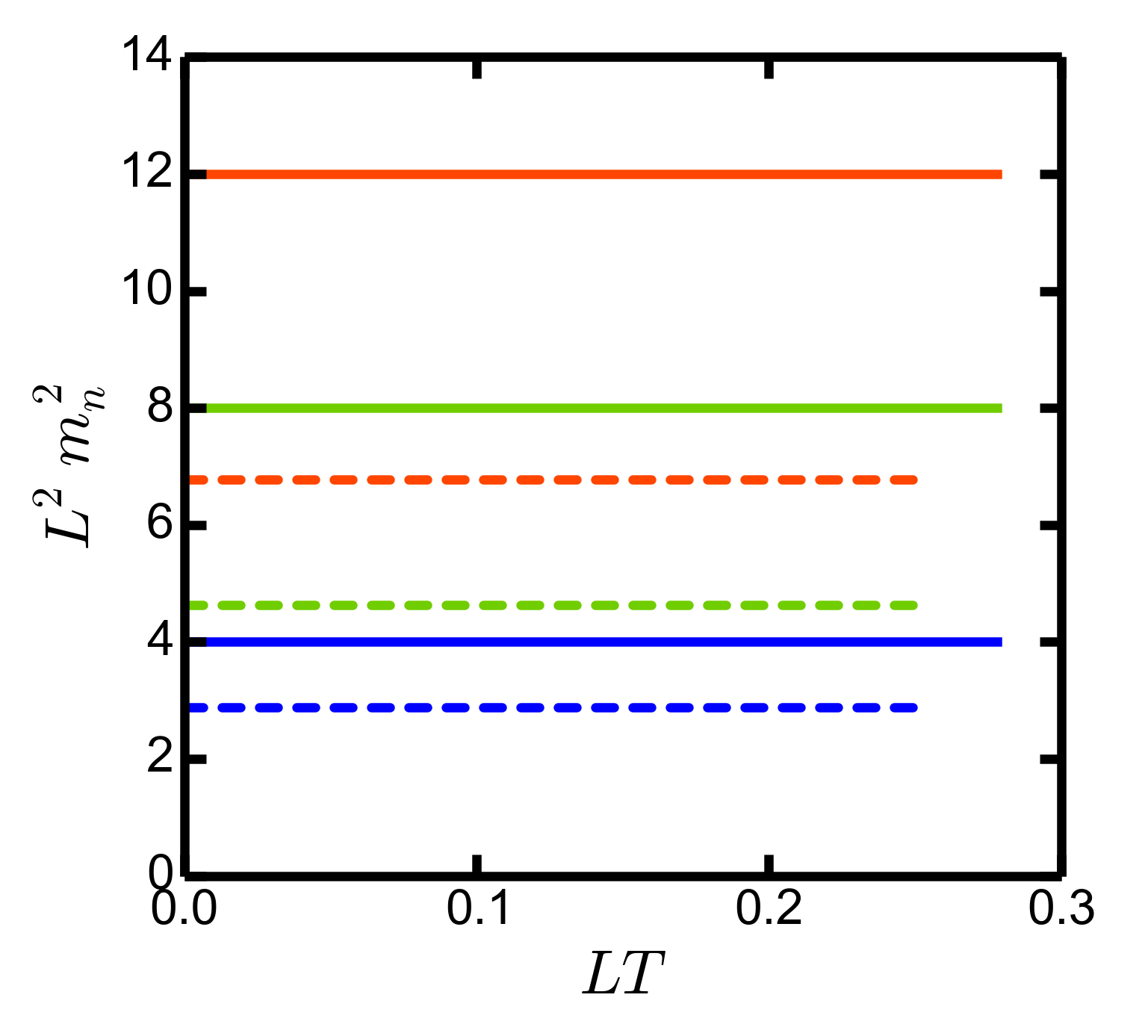}
      \caption{Vector meson masses squared $L^2m_n^2$ of the first three states (color code as in Figure~\ref{abb5}) as a function of the temperature for the parameter $p=2$ (solid) and $p=1$ 
(dashed). The masses are constant during the thermal gas phase and all excitations disappear at $T_{\rm c}$, where the black hole solution begins to apply.} \label{abb6}}
     \end{figure}

To get an idea on the related scale, the left-hand plot of Figure~\ref{abb5} shows the first three states, that is $L^2m_n^2$ as a function of $p$ for $n=0$ (ground state) and $n=1$, 2 
(first two radial excitations). For comparison, the middle plot of Figure~\ref{abb5} displays $LT_{\min}$ also as a function of $p$. Both figures can be combined to $T_{\min}/m_n$ as a function of 
$p$ 
(right-hand plot of Figure~\ref{abb5}). Having in mind applications to QCD and identifying $T_{\min}$ with 155~MeV (see above) and $m_0$ with the $\rho$ meson ground state mass of 770~MeV, one 
arrives 
at $T_{\min}/m_0 \approx 0.20$, i.e.~a value not too far from the range of values shown in the right-hand plot of Figure~\ref{abb5}. 
However, 2+1 flavor QCD with physical quark masses does not provide a first-order phase transition. Insofar, the present set-up is more appropriate for 2+1 flavor QCD in the chiral limit, which in 
fact enjoys a first-order phase transition \cite{philipsen2018, ding}, but detailed information on thermodynamic quantities as well as the vector meson spectrum is lacking (cf.~\cite{Karsch} for a 
search for the delineation curve in the Columbia plot where cross-over and first-order phase transitions touch each other; very preliminary first estimates \cite{schmidt} point to a ratio of $T_{\rm 
c}/m_0$ in the same order of magnitude as for the case of physical quark masses); $T_{\rm c}$ becomes in the chiral limit $\approx 132$~MeV \cite{ding}. \\
The temperature dependence of $U_T$ (not shown) is such to cause the instantaneous disappearance (melting) of vector meson states at $T_{\rm c} \approx T_{\min}$ (see Figure~\ref{abb6}). Denoting the 
disappearance temperature by $T_{\rm dis}$ and choosing $m_0=m_{\rho}$ as scale, Figure~\ref{abb6} translates into $T_{\rm dis}=m_{\rho} \frac{LT_{\rm dis}}{Lm_0}$, meaning $T_{\rm dis}=116$~MeV for 
$LT_{\rm dis}=0.27\,(0.3)$ and $Tm_0=1.7\,(2)$. That value of $T_{\rm dis}$ does not apply to 2+1 flavor QCD with physical quark masses, since, as stressed above, the present scenario is more 
suitable for the chiral limit where a first-order phase transition occurs.

\section{Summary and discussion} \label{sec5}
We focus here on the QCD relevant cross-over transition and its discrimination against first-order and second-order transitions. Obviously, more complicated structures are
possible, e.g.~a sequence or nested first-order transitions for functions $T(z_H)$ with multiple local minima (see \cite{kir1} for the case of
a double transition). These require further shaping of the dilaton
potential. This can be easily done by combining the elements of our systematics presented in this paper: an extreme point of $\Dp_{\Phi} V/V$ generates an inflection point of $T(z_H)$ which points to 
a cross-over or a second-order phase transition (if it is a horizontal turning point) or to a first-order phase transition if the temperature exhibits additional extreme points which can be controlled 
by the altitude of $\Dp_{\Phi} V/V$.\\
We did not touch such issues as good and bad curvature singularities \cite{singular}, adding further (e.g.~charged scalar) fields which can bridge to order parameters and/or condensation 
\cite{cubro}, larger classes of dilaton potentials (e.g.~Liouville potentials or linear combination thereof \cite{kirfr}) and fluctuations. \\
Unfortunately, the Einstein gravity - dilaton model seems to be not flexible enough to allow simultaneously for a cross-over and probe meson states because the existence of the 
latter ones requires a minimum in $T(z_H)$. The obvious idea to construct a dilaton potential such that $T(z_H)$ has a minimum at a horizon $z_H^*$ with $T(z_H^*)$ being small (to ensure the existence 
of the mesons) and a cross-over at the QCD critical temperature of about 155~MeV does not work very well, since all probe mesons states disappear already at the minimum temperature $T(z_H^*)$. 
\blau{However, keeping the background, as determined in Section \ref{sec3}, a refinement of
the action (\ref{mesonwirkung}) has the capability of describing properly (i) probe vector mesons
at low temperatures and (ii) the pattern of meson melting at high temperatures,
consistent with lattice QCD. The details will be reported elsewhere.}\\
A first step further on the road to a fully consistent approach could be the consideration of a $U(1)$ Maxwell type gauge field. Such a field has been used to address the question of the behavior  
of probe vector mesons in relation to the thermodynamics: $\rho$ mesons are described through the $U(1)$ gauge field (see (\ref{mesonwirkung})) and putting together the actions (\ref{wirkung}) and 
(\ref{mesonwirkung}) would yield a model with full back reaction from the mesons to the gravity background. \\
Since QCD thermodynamics is not driven by vector mesons alone, another step is adding systematically flavor, e.g.~by including the pseudo-scalar and scalar sectors via the bulk 
fundamental fields and its vacuum expectation values. Some works point directly in this direction: the authors of \cite{bk2} introduce a second scalar field (glue ball field) and solve the field 
equations for the case $T=0$; many other investigate the behavior of hadron species in a given background without back reaction (see e.g.~\cite{col12, col09, cui11, cui13, lee09, fuj09, mir09, 
ich2016, csaki}). \cite{yamaguchi, fang} give a study of phase transitions in relation to a flavor containing model with given metric background. Back reactions are accounted for in the 
V-QCD model class pioneered in \cite{vqcd}, where the flavor sector supplements the gluon (dilaton) sector, thus catching many desired features in relation to QCD, up to the equation of state 
adapted to the 2+1 flavor case for $T > T_{\rm c}$ (cf.~\cite{vqcd2} and further references therein). Bringing the characteristic features of the 
mentioned works together would be an improvement. The Holy Grail would be a model with parameters steering quark masses and condensates for different flavors separately with proper hadron spectra in 
all (scalar, pseudo-scalar, vector, axial-vector, tensor and axial-tensor) sectors.\\
All mentioned extensions point to the leading question, which framework is needed to have a QCD consistent thermodynamics and proper in-medium modifications of the hadron species. 
However, increasing the variety of a model means increasing its complexity and requires much follow-up work.

\begin{acknowledgments}
The authors gratefully acknowledge useful discussions with M. Ammon, J. Erdmenger, M. Huang, M. Kaminski, J. I. Kapusta, J. Knaute and J. Noronha. The work of RZ is supported by 
Studienstiftung des deutschen Volkes. 
\end{acknowledgments}

\begin{appendix}

\section{Derivation of (\ref{T1})} \label{anhang_A}
We use $f'(z_H,z_H) = -4\pi T(z_H)$ and $f(z_H,z_H)=0$ to evaluate (\ref{fgl3}) and (\ref{fgl4}) at $z=z_H$ which imply
  \begin{eqnarray}
   T(z_H) &=&\frac1{6\pi}\frac{e^{A(z_H,z_H)}}{A'(z_H,z_H)} V(\Phi(z_H,z_H)), \label{eins} \\
   T(z_H) &=& -\frac1{4\pi}\frac{e^{A(z_H,z_H)}}{\Phi'(z_H,z_H)} \Dp_{\Phi} V(\Phi(z_H,z_H)) ,\label{zwei}
  \end{eqnarray}
respectively. Differentiating (\ref{eins}) w.r.t.~$z_H$ yields
  \begin{equation}
   \frac1T \frac{\dd T}{\dd z_H} = A' +\frac{\Dp A}{\Dp z_H} -\frac1{A'} \li(A'' +\frac{\Dp A'}{\Dp z_H}  \ri) + \frac{\Dp_{\Phi} V}{V} \frac{\dd \Phi}{\dd z_H}, \label{zweia}
  \end{equation}
where all functions are to be taken at $(z_H,z_H)$. Equating (\ref{eins}) and (\ref{zwei}) yields $A'$ at $z=z_H$ as
  \begin{equation}
   A' \mid_{z=z_H}  = -\frac23 \Phi' \frac{V}{\Dp_{\Phi} V} \mid_{z=z_H}.   \label{drei}
  \end{equation}
By inserting (\ref{drei}) in (\ref{zweia}) and eliminating $A''$ via (\ref{fgl2}) we find 
  \begin{equation}
   \frac1{\Phi' T} \frac{\dd T}{\dd z_H} = \frac12 \frac{\Dp_{\Phi} V}{V} -\frac13 \frac{V}{\Dp_{\Phi} V} + \frac1{\Phi'} \frac{\Dp A}{\Dp z_H}+ \frac1{\Phi'} \li(\frac3{2 \Phi'} \frac{\Dp A'}{\Dp z_H} + \frac{\Dp \Phi}{\Dp z_H}\ri) \frac{\Dp_{\Phi} V}{V} \label{vier}
  \end{equation}
at $z=z_H$. This leads directly to (\ref{T1}). The next step is to solve the field equation (\ref{fgl1}):
  \begin{equation}
   f(z,z_H) = 1-\frac{h(z,z_H)}{h(z_H,z_H)},
  \end{equation}
where $h(z,z_H) := \int \nolimits_0^{z} \exp(-3/2 A(\tilde z,z_H)) \, \dd \tilde z$. This solution is well defined and can be employed to compute the temperature a for third time:
  \begin{equation}
   T(z_H) = \frac1{4\pi} \frac{e^{-\frac32 A(z_H,z_H)}}{h(z_H,z_H)}.
  \end{equation}
After differentiating (\ref{eins}) w.r.t.~$z_H$ and some manipulations we end at
  \begin{equation}
   \frac1{T} \frac{\dd T}{\dd z_H} +4\pi T = -\frac32(A'+\frac{\Dp A}{\Dp z_H})\mid_{z=z_H}  +\frac3{2h} \int \limits_0^{z_H} \frac{\Dp A}{\Dp z_H}(\tilde z,z_H) e^{-\frac32 A(\tilde z,z_H)} \, \dd \tilde z \label{sieben}
  \end{equation}
which we will use in Appendix \ref{anhang_B}.

\section{Derivation of (\ref{final})} \label{anhang_B}

We start with (\ref{u0}). To relate $U_0$ with $T$, we evaluate (\ref{eins}, \ref{zwei}, \ref{vier}, \ref{sieben}) in leading order, i.e.~neglecting $a$ and $\varphi$:
  \begin{eqnarray}
   T &=& \frac1{6\pi}\frac{e^{A_0}}{A_0'} V(\Phi_0), \\
   T &=& -\frac1{4\pi}\frac{e^{A_0}}{\Phi_0'} \Dp_{\Phi} V(\Phi_0), \\
   \frac1{\Phi_0' T} \frac{\dd T}{\dd z_H} &=& \frac12 \frac{\Dp_{\Phi} V}{V} -\frac13 \frac{V}{\Dp_{\Phi} V}, \\
   \frac1{T} \frac{\dd T}{\dd z_H} +4\pi T &=& -\frac32A_0' .
  \end{eqnarray}
These four equations allow for eliminating all $A$'s and $\Phi$'s in (\ref{u0}). Finally we arrive at (\ref{final}).

\section{Towards a systematic dilaton potential expansion analysis}

  \begin{figure}
   \cen{\includegraphics[scale=.48]{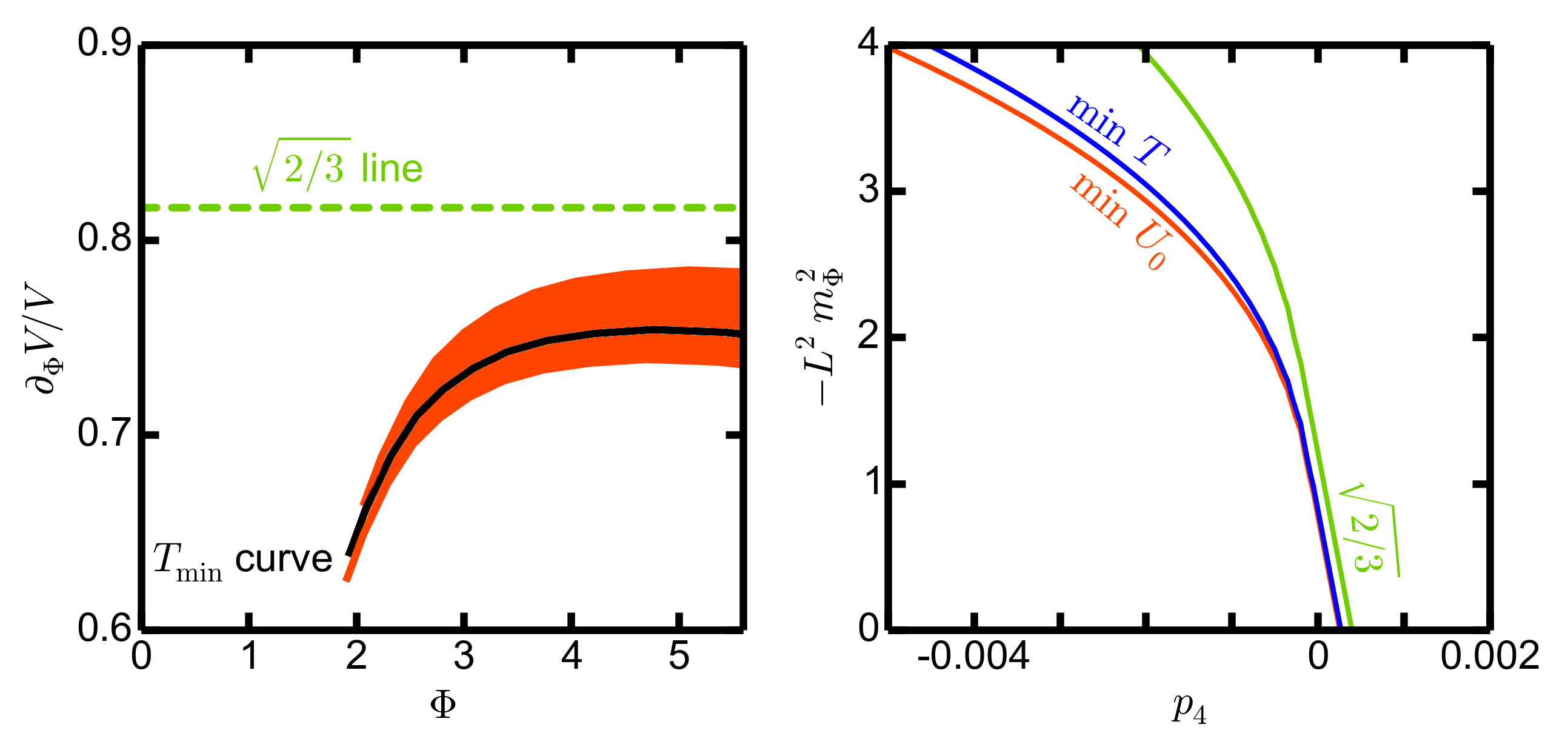}
   \caption{Left: $T_{\min}$ curve (solid black) \blau{generated by} the potential (\ref{hase}) \blau{under variation of $p_2$ at} $p_{i \geq 3}=0$, analogue to Figure~\ref{abb3}. By varying a second parameter 
(here $p_3= -0.01 \cdots 0.01$ \blau{and  $p_{i \geq 4}=0$}), the 
$T_{\min}$ curve becomes a strip (red area), where a negative (positive) value of $p_3$ belongs to the lower (upper) part, since increasing $p_3$ means increasing the slope of $\Dp_{\Phi} V/V$ and 
therefore increasing also the difference between Gubser's adiabatic criterion and the $T_{\min}$ curve. If $\Dp_{\Phi} V/V$ as a function of $\Phi$ with given parameter set $\vec p$ has a section within 
the corresponding $T_{\min}$ area (here only displayed \blau{for} a $p_3$ interval) 
then a first-order or HP phase transition is facilitated.
Right: Border lines of parameter regions in $-L^2 m_{\Phi}^2=p_2/24$ vs. $p_4$ space, where either $T(z_H)$ has a minimum (right to blue curve), or $U_0(z)$ has a minimum (right to red curve), or $\Dp_{\Phi} V/V$ is greater than $\sqrt{2/3}$ (right to green curve) for the dilaton potential (\ref{hase}) with $p_3=0$ and $p_{i \geq 5}=0$; if $m_{\Phi}$ and therefore the slope of $\Dp_{\Phi} V/V$ increase, the difference between Gubser's adiabatic criterion and the $T_{\min}$ curve becomes larger.} \label{abb7}}
  \end{figure}
  
    \begin{figure}
   \cen{\includegraphics[scale=.48]{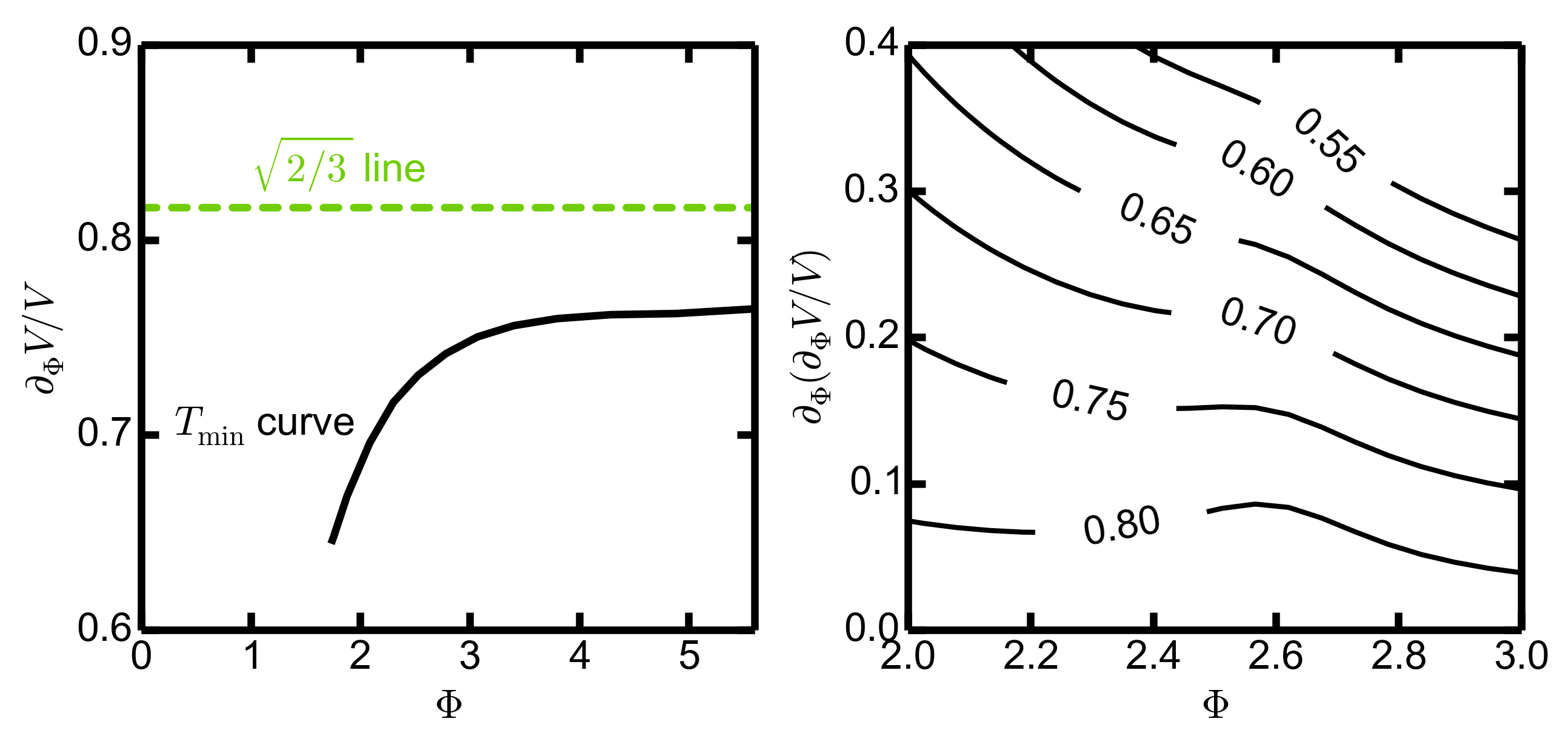}
   \caption{\blau{Left: $T_{\min}$ curve (solid black) generated by the potential (\ref{polyv}) under variation of $\phi_2$ at $\phi_4=0.23$ and $\phi_{i \geq 6} =0$, analogue to Figure~\ref{abb3}. 
Right: contour plot of the minimum value $(\Dp_{\Phi} V/V)^*$ over the plane $\Phi$ vs.~$\Dp_{\Phi}(\Dp_{\Phi} V/V)$ for which $T(\Phi_H)$ exhibits a minimum: If $(\Dp_{\Phi} V/V)$ exceeds this value 
(depending on its slope and the position) the temperature exhibits a minimum which is independent of the concrete set of parameters.}\label{abb7_neu}} } 
  \end{figure}

Another useful form of the dilaton potential is
  \begin{equation}
   -L^2V(\Phi) = 12 \exp \li(\sum \nolimits_{i=2} p_{i} \Phi^{i} \ri), \label{hase}
  \end{equation}
since $\Dp_{\Phi} V/V$ runs over all BF permitted and AdS conform polynomials if $\vec p=(p_2,p_3,\cdots)$ runs over all vectors. The case $p_{i \geq 3}=0$ characterizes the leading order (straight lines) in the spirit of an expansion of $\Dp_{\Phi} V/V$ in powers of $\Phi$.
The left panel in Figure~\ref{abb7} shows the influence of the varying $p_3$-depending term (red strip) on the $T_{\min}$ curve (black curve) which is generated with running $p_2$ analogously to 
Figure~\ref{abb3}. The parameter regions in the $-L^2m_{\Phi}^2$ vs. $p_4$ plane, where $T(z_H)$ (blue curve) or $U_0(z)$ (red curve) exhibits a minimum, as well as the area, where $\Dp_{\Phi}V/V$ is 
greater than $\sqrt{2/3}$ (green curve), are also shown (see right panel and compare with Figure~\ref{abb3a}). In such a manner one can study, piece by piece, the impact of the individual terms in 
(\ref{hase}) on the issue of phase structure and capabilities to permit vector meson modes in the probe limit. \\
\blau{We complement the ansatz (\ref{hase}) by a purely polynomial form of $V(\Phi)$ in the spirit of a small-$\Phi$ expansion:\footnote{\blau{We thank the anonymous Referee for that suggestion.}}
  \begin{equation}
   -L^2V(\Phi) = 12 + \sum \limits_{i=1} \phi_{2i} \Phi^{2i} \label{polyv}
  \end{equation}
and vary the parameters $\phi_2, \, \phi_4$ and $\phi_6$.
A $T_{\min}$ curve is exhibited in Figure \ref{abb7_neu}-left panel (cf.\ Figures \ref{abb3} and \ref{abb7} for other potential ans\"atze).
In addition, the right panel of Figure \ref{abb7_neu} displays a contour plot of the quantity $(\Dp_{\Phi}V/V)^*$ over the plane spanned by the coordinates $\Phi$ and $\Dp_{\Phi}(\Dp_{\Phi}V/V)$.
The meaning of $(\Dp_{\Phi}V/V)^*$ is as follows: that value is the minimum at which the corresponding curve $T(\Phi_H)$ acquires a local minimum, thus turning the smooth (cross-over) thermodynamic 
behavior into a first-order or HP transition. Note that the quantity $\Dp_{\Phi}(\Dp_{\Phi}V/V)$ is the slope of $\Dp_{\Phi}V/V$ and so Figure \ref{abb7_neu} could be understood as some kind of 
Legendre transform parametrizing a function by its derivative. In essence, the shown minimum value of $\Dp_{\Phi}V/V$ is needed 
to produce at least a local minimum of the temperature as a function of $z_H$ or $\Phi_H$. The particular value of such an analysis is that the minimum value of $\Dp_{\Phi}V/V$ depends mainly on 
$\Phi$ and the slope of the quantity $\Dp_{\Phi}V/V$ and not on the concrete choice of parameters which leads to the combination of $\Phi$ and $\Dp_{\Phi}(\Dp_{\Phi}V/V)$.\\
We emphasize again that, w.r.t.\ applications for QCD$_{2+1}$(phys.), parameter regions which facilitate a first-order order or HP transitions must be avoided. In fact, there are parameter regions 
which allow the desired cross-over as well. Insofar, (\ref{polyv}) can provide a guidance for such a goal upon a small-$\Phi$ expansion of specific ans\"atze of the dilaton 
potential.}

\end{appendix}

\end{document}